\pdfoutput=1
\PassOptionsToPackage{table}{xcolor}
\documentclass[acmsmall,screen,nonacm]{acmart}
\usepackage{tabularx}
\usepackage{paper5_style}
\newcommand{\ActiveThreshold}{0.05}
\newcommand{\AgreeSpec}{98}

\newcommand{\AtlasRateAlcohol}{7.4}

\newcommand{\AtlasRateUnbelted}{1.7}
\newcommand{\AtlasRateWrongWay}{6.7}

\newcommand{\AucFlagHardSame}{0.865}
\newcommand{\AucFlagKumo}{0.865}
\newcommand{\AucFlagKumoLargeTen}{0.853}

\newcommand{\AucFlagKumoMediumTen}{0.847}

\newcommand{\AucFlagKumoTen}{0.849}

\newcommand{\AucFlagKumoThousand}{0.792}
\newcommand{\AucFlagPFNNew}{0.872}

\newcommand{\AucFlagSoft}{0.856}

\newcommand{\AucLine}{0.85}

\newcommand{\AucSixteenPhone}{0.72}

\newcommand{\AucSixteenUnbelted}{0.64}

\newcommand{\BRecal}{200}
\newcommand{\BenchCtx}{60,000}
\newcommand{\BenchDisc}{three}
\newcommand{\BenchFlags}{eight}
\newcommand{\BenchTest}{30,000}

\newcommand{\BreakevenMin}{2.44}

\newcommand{\CfAucBooster}{0.881}
\newcommand{\CfAucKumo}{0.890}

\newcommand{\CfNTargets}{18}
\newcommand{\CodedAlcohol}{73,809}

\newcommand{\CodedMedical}{16,193}
\newcommand{\CodedPhone}{7,340}
\newcommand{\CodedUnbelted}{57,657}

\newcommand{\CompanionCoded}{195,857}

\newcommand{\ConRuralAnimal}{3.0}
\newcommand{\ConRuralAnimalHi}{3.4}
\newcommand{\ConRuralAnimalLo}{2.6}

\newcommand{\CondShareAnimal}{15}

\newcommand{\CondShareWrongWay}{69}

\newcommand{\ConfFalseUnbeltedN}{13}

\newcommand{\CostFullCoding}{773}

\newcommand{\CountyMinCrashes}{10,000}
\newcommand{\CovAxisFloor}{75}
\newcommand{\CovMPD}{0.972}
\newcommand{\CovSeq}{0.952}
\newcommand{\CovWaveOne}{0.948}
\newcommand{\CtxRows}{60,000}

\newcommand{\EceCatBoost}{0.0023}

\newcommand{\EcePFNNew}{0.0020}

\newcommand{\EstAlcohol}{65,971}

\newcommand{\EstHitRun}{143,701}

\newcommand{\EstMedical}{28,345}

\newcommand{\EstPhone}{15,074}
\newcommand{\EstTransported}{224,996}
\newcommand{\EstUnbelted}{9,617}
\newcommand{\EstWitness}{135,504}

\newcommand{\FatalShareOutU}{0.17}
\newcommand{\FatalShareU}{0.64}

\newcommand{\FullGainAlcohol}{2}

\newcommand{\GammaHi}{4.50}
\newcommand{\GammaLo}{0.37}
\newcommand{\GammaShift}{2.00}

\newcommand{\GpuKumoLargeTen}{45.4}

\newcommand{\GpuKumoMediumThirty}{66.6}

\newcommand{\HumAucBoosterOne}{0.874}
\newcommand{\HumAucBoosterTwo}{0.899}

\newcommand{\HumAucKumoOne}{0.913}
\newcommand{\HumAucKumoTwo}{0.899}

\newcommand{\HumanCostUsd}{0.60}

\newcommand{\JointBreakeven}{12.68}

\newcommand{\JointRmsFull}{21.2}
\newcommand{\JointRmsNow}{22.0}
\newcommand{\JointRmsPlus}{16.2}
\newcommand{\KappaNonspecHi}{0.57}
\newcommand{\KappaNonspecLo}{0.50}
\newcommand{\KappaSpec}{0.93}
\newcommand{\Kfolds}{5}

\newcommand{\KumoAucFatigue}{0.95}
\newcommand{\KumoAucHitRun}{0.93}

\newcommand{\KumoAucPhone}{0.89}
\newcommand{\KumoAucTransported}{0.91}
\newcommand{\KumoAucUnbelted}{0.77}
\newcommand{\KumoAucWitness}{0.78}
\newcommand{\KumoAucWrongWay}{0.81}

\newcommand{\LabelRound}{400}

\newcommand{\LiftOneAlcohol}{42}
\newcommand{\LiftOneAnimal}{58}

\newcommand{\LiftOneHydroplane}{41}

\newcommand{\LiftOneMedical}{35}

\newcommand{\LiftOneUnbelted}{7}
\newcommand{\LiftOneWrongWay}{11}

\newcommand{\LowpDefect}{41}
\newcommand{\LowpOtherMax}{0.6}
\newcommand{\LowpWrongWay}{4}
\newcommand{\MapSpreadDefect}{21}
\newcommand{\MapSpreadMax}{8}
\newcommand{\NAll}{5,601,890}
\newcommand{\NBenchTargets}{eleven}

\newcommand{\NBetterPFNNew}{five}

\newcommand{\NEmptyNarr}{99,332}
\newcommand{\NFatalOutU}{997}
\newcommand{\NFeatures}{97}
\newcommand{\NFeaturesCat}{26}
\newcommand{\NLongWithCoords}{16}
\newcommand{\NNonGeo}{484,462}

\newcommand{\NSoft}{eight}
\newcommand{\NSpecShared}{148}
\newcommand{\NU}{5,018,080}
\newcommand{\NWorseCatBoost}{six}

\newcommand{\NWorsePFNNew}{none}
\newcommand{\NWorsePFNOld}{four}
\newcommand{\NWorseTabICL}{seven}

\newcommand{\NarrOnlyAnimal}{15}

\newcommand{\NarrOnlyFatigue}{35}

\newcommand{\NarrOnlyUnbelted}{80}
\newcommand{\NarrOnlyWrongWay}{75}
\newcommand{\NonspecYesShare}{59}

\newcommand{\OneHwPhone}{16.8}

\newcommand{\PassHoursPFN}{2.8}
\newcommand{\PassMinKumo}{11}

\newcommand{\PiCap}{0.95}
\newcommand{\PiTwoMax}{0.68}
\newcommand{\PiTwoMin}{0.0057}

\newcommand{\PlusGainAlcohol}{28}

\newcommand{\PrecOneAlcohol}{31.6}

\newcommand{\RandPrecAlcohol}{0.76}

\newcommand{\RecalShareAlcohol}{95}

\newcommand{\RecalShareWrongWay}{99}

\newcommand{\RetroRatio}{1.29}
\newcommand{\RetroRatioHi}{1.32}
\newcommand{\RetroRatioLo}{1.13}

\newcommand{\RhoSqHardSame}{0.375}
\newcommand{\RhoSqKumo}{0.279}

\newcommand{\RhoSqPFNNew}{0.286}

\newcommand{\RhoSqSoft}{0.376}

\newcommand{\RoneAgree}{99}
\newcommand{\RoneDisputed}{48}

\newcommand{\RtwoCoded}{600}
\newcommand{\RtwoCompletionCC}{14}
\newcommand{\RtwoConsConsistent}{thirteen}
\newcommand{\RtwoConsistent}{fifteen}

\newcommand{\RtwoDrawn}{757}

\newcommand{\RtwoFrame}{149,600}

\newcommand{\RtwoJudg}{2,397}
\newcommand{\RtwoMedical}{28,825}
\newcommand{\RtwoMedicalHi}{31,816}
\newcommand{\RtwoMedicalLo}{25,834}
\newcommand{\RtwoNAbove}{five}
\newcommand{\RtwoNInformative}{eight}
\newcommand{\RtwoNarr}{427}

\newcommand{\RtwoNonspecAlcohol}{11,660}

\newcommand{\RtwoPhone}{16,211}
\newcommand{\RtwoPhoneHi}{19,371}
\newcommand{\RtwoPhoneLo}{13,050}
\newcommand{\RtwoPiMax}{0.34}

\newcommand{\RtwoWithheld}{112}
\newcommand{\RtwoWithheldPct}{15}

\newcommand{\ScaleAlphaMax}{5.74}
\newcommand{\ScaleAlphaMedian}{1.06}
\newcommand{\ScaleAlphaMin}{\ensuremath{-}1.95}
\newcommand{\SchemaQuestions}{27}

\newcommand{\ShareGeoU}{89.6}

\newcommand{\ShiftMaxChange}{9}
\newcommand{\ShiftMaxFactor}{wrong-way driving}
\newcommand{\SimDesigns}{500}
\newcommand{\SimNarrowing}{30}
\newcommand{\SlopeCatBoost}{1.01}

\newcommand{\SlopeKumo}{1.01}

\newcommand{\SlopePFNNew}{1.02}

\newcommand{\SlopeTabICL}{0.90}

\newcommand{\SpeedRatioPFN}{15}

\newcommand{\TableGainMax}{64}
\newcommand{\TableGainMaxFactor}{animals}
\newcommand{\TableGainMedian}{19}

\newcommand{\TableGainTotalMedian}{2}

\newcommand{\TempCtxRows}{12,000}

\newcommand{\TempPhoneMatchedTwentyOne}{0.889}

\newcommand{\TempPhoneMismatchedTwentyOne}{0.831}

\newcommand{\TwoHwPhone}{16.0}

\newcommand{\WaveTwoFrame}{4,868,080}

\newcommand{\cOne}{0.65}
\newcommand{\nGoldJudg}{2,309}
\newcommand{\nGoldNarr}{400}
\newcommand{\nOne}{150,000}
\newcommand{\nTwoDraws}{79,510}
\newcommand{\nTwoExpControl}{20,000}
\newcommand{\nTwoExpScore}{60,000}
\newcommand{\nTwoNew}{78,267}
\newcommand{\nTwoReuse}{1,243}

\begin{document}
\setcopyright{cc}
\setcctype{by}
\copyrightyear{2026}
\acmYear{2026}
\acmDOI{XXXXXXX.XXXXXXX}
\acmJournal{TIST}
\acmVolume{0}
\acmNumber{0}
\acmArticle{0}
\acmMonth{1}

\title[Estimating Uncoded Crash Factors]{Estimating Uncoded Crash Factors with Tabular Foundation and System One Models: Kumo Tabular and Jev}

\author{Amir Rafe, Ph.D.}
\orcid{0000-0002-4089-2088}
\affiliation{\institution{Texas State University}\city{San Marcos}\country{USA}}
\email{amir.rafe@txstate.edu}
\author{Subasish Das, Ph.D.}
\orcid{0000-0002-1671-2753}
\affiliation{\institution{Texas State University}\city{San Marcos}\country{USA}}
\email{subasish@txstate.edu}
\renewcommand{\shortauthors}{Rafe and Das}

\begin{abstract}
Road safety programs count the coded fields of police crash records, while the officer's narrative, which often records factors the fields omit, is rarely read. A safety office thus cannot tell how much its counts miss or where to review. This study develops and evaluates a system that joins both views of the \NAll{} Texas crashes from 2017 to 2025 into population estimates with stated validity. An in-context tabular foundation model, Kumo Tabular, reads the coded record of every crash, a calibrated System One model, Jev, reads the narratives of two probability samples, and human judgments recalibrate its probabilities. A multiwave predict-then-debias estimator joins the three tiers, and a second human tier drawn with recorded probabilities checks the estimates by design. For hydroplaning, medical episodes, fatigue, animals, and phone use, the narrative documents more injury crashes than the coded field, \EstPhone{} against \CodedPhone{} for phone use, and the human check agrees with all \RtwoConsistent{} estimates within its margin. A re-read list ranked by Kumo Tabular finds confirmed discordance \LiftOneUnbelted{} to \LiftOneAnimal{} times as often as random reading. At the planning cost of human coding, one further round of human judgments would cut the root mean square relative half-width from \JointRmsNow{} to \JointRmsPlus{} percent, against \JointRmsFull{} for reading every narrative. Two calibrated readers of different views, joined by a sampling design, give a safety office counts, a discordance map, a validated re-read list, and a reading budget, with Kumo Tabular reading the table at \SpeedRatioPFN{} times the speed of TabPFN 3.5.

\end{abstract}


\maketitle

\section{Introduction}\label{sec:intro}

Road safety programs decide where to act from police crash records, and the decisions are only as good as what those records hold. Every state collects a crash report for each police-reported crash, codes it into fields that feed safety analysis, countermeasure selection, and federal reporting, and keeps the investigating officer's written account beside the fields. The coded record is what programs count, and the narrative is where the officer writes that a driver fell asleep, had a seizure, was looking at a phone, or swerved for a deer. Machine learning has recently reached the point where both parts of the record can be read by models at the scale of a state, through tabular foundation models that read structured records without training and language-based decision models that read text and return probabilities. Whether these readers can turn the full record into numbers that a safety office can trust is therefore a timely question for intelligent transportation systems.

The specific problem is that the two views of a crash record are read separately and almost never measured against each other at the scale at which decisions are made. A coded field records what the form asks and what the officer chose to code, while the narrative records what the officer observed or was told, so the same crash can carry a factor in one view and not in the other. A safety office that counts only the coded field cannot know how much it misses, and a narrative read by a model on a sample cannot become a statewide count without a sampling design and a human reference that defines what a narrative states. The problem sits at the meeting point of crash-data quality, text and tabular machine learning, and survey inference with model predictions.

Each of these lines of work supplies part of what the problem needs and leaves the rest open. Studies of crash-data quality document that police records under-report casualties and that factors such as distraction, fatigue, and impairment depend on what an officer can establish at the scene~\cite{elvik1999incomplete,janstrup2016understanding,imprialou2019crash,filtness2017sleep,sjogren1997comparison}. They measure that gap against hospital or investigator records on samples, not against the narrative of the same report. Narrative studies, several of them with large language models, compare narratives with official coding~\cite{fitzpatrick2017investigation,hosseini2023application,bharati2026benchmarking,arteaga2025large,das2026revealing}, and they report agreement on a sample, which is not a population count and says nothing about where the disagreement lies. Prediction-powered inference and its multiwave extension give valid intervals from model predictions and a smaller set of trusted measurements~\cite{angelopoulos2023prediction,angelopoulos2026ppi,kluger2026estimation}, but they use a single proxy and have not been applied to crash records. Tabular models read the coded record~\cite{somvanshi2025tabular,rafe2024comparative}, yet only to predict an outcome, not as a reader inside an inference design. Earlier studies of Texas narratives counted narrative-derived factors~\cite{rafe2026calibrated,rafe2026counting}, but from the narrative view alone.

Despite these advances, no study has linked the two views of the same crash record under one sampling design. No study has turned the disagreement between a coded field and the narrative into statewide counts with intervals or located that disagreement by road system, setting, and county. None has validated a list of records worth reading again on a probability sample or computed how a further reading budget should be split between machine reads and human judgments. These are the quantities a safety office needs to decide which fields to revise, which records to review, and where to spend the next reading budget. Two models released in 2026 make these quantities computable at the scale of a state. Kumo Tabular is a tabular foundation model with class probabilities and a license for commercial use that had no independent evaluation at the time of this study~\cite{qu2026nvidia}, and Jev is a calibrated System One decision model for text~\cite{typesafe2026launch,typesafe2026systemone}.

The aim of this paper is to develop and evaluate a system that joins the two views of the crash record for the Texas crashes of 2017 to 2025 and turns them into population estimates whose validity is stated tier by tier. The paper addresses five research questions about the system and its outputs. RQ1 asks how much of what each narrative states is predictable from the coded record and how much is surplus that only the narrative carries. RQ2 asks how many injury crashes in Texas each year have a narrative that states each factor, with intervals whose validity is stated tier by tier. RQ3 asks where narratives and coded fields disagree after human confirmation, by year, road system, population group, county, setting, severity, and road class. RQ4 asks whether a re-read list ranked by the table reader finds confirmed discordance more often than random reading, and how a future budget should be split between narrative reads and human judgments. RQ5 asks how Kumo Tabular compares with TabPFN, TabICLv2, and a trained booster as a calibrated in-context reader of a crash table.

The novelty of the paper lies in composing two decision-model readers of different views and a human tier into one inference design. Kumo Tabular reads the coded record of every crash, Jev reads the narratives of two probability samples, human judgments recalibrate Jev, and the multiwave predict-then-debias estimator of Kluger and Bates joins the three tiers into estimates whose intervals include the uncertainty of the human recalibration. A second human tier, drawn with recorded probabilities, checks the estimates by design. The main contribution is a measured account of what each view of the crash record carries, delivered as population counts, a map of confirmed discordance, a validated re-read list, and a reading budget, together with an independent evaluation of Kumo Tabular on a real population. Every quantity is an estimate of a population total, an agreement rate, or a benchmark score, and the paper makes no causal claim.

Section~\ref{sec:background} reviews the four lines of research that the paper joins. Section~\ref{sec:data} describes the data and the phases of the design, and Section~\ref{sec:methods} describes the readers, the estimator, the human check, and the outputs. Section~\ref{sec:results} answers the research questions, Section~\ref{sec:discussion} compares the findings with prior studies, and Section~\ref{sec:conclusion} concludes.

\section{Background}\label{sec:background}

Police crash data are known to record less than what happened, and the gap varies with the factor and with the evidence an officer can collect. A meta-analysis of studies in thirteen countries found that official statistics capture fatalities almost completely and slight injuries far less often~\cite{elvik1999incomplete}, and linked police and hospital records in Denmark showed that reporting depends on road user type and injury level~\cite{janstrup2016understanding}. A review of crash data quality ranked inaccurate contributory factors among the most serious problems for safety research~\cite{imprialou2019crash}. Factor-level studies point the same way. Police reports of belt use had high sensitivity but low specificity against crash investigators~\cite{schiff2004comparison}, and police assessments of alcohol and drugs missed a large share of positive blood tests~\cite{sjogren1997comparison}. A proxy definition also found nearly three times as many sleep-related crash involvements as police reports did~\cite{filtness2017sleep}. Part of this gap is procedural rather than erroneous. The Texas instructions to police code only one alcohol result, place portable breath test results in the narrative alone, and treat contributing factors as the investigator's opinion~\cite{txdot2025instructions}, so a coded field can lag the narrative without being wrong.

Reading the narrative has moved from keyword rules through supervised classifiers to language models. Injury surveillance used Bayesian and machine-learning autocoders with human review of uncertain cases~\cite{marucciwellman2017classifying}, and transformer classifiers coded injury types in hundreds of thousands of crash narratives~\cite{oliaee2023using}. A blind review of crash narratives found that about half of the crashes designated as speeding-related described speeding~\cite{fitzpatrick2017investigation}. Language models now read narratives to compare them with coded fields. A benchmark of six frontier models against Arkansas fatal-crash coding found that a keyword baseline matched the best model and that differences across attributes exceeded differences across models~\cite{bharati2026benchmarking}. Other studies found under-reported alcohol involvement~\cite{arteaga2025large} and discrepancies in both directions between narratives and coded impaired-driving fields~\cite{das2026revealing}, and an evaluation found that higher accuracy against the coded label can mean lower agreement with experts~\cite{bhagat2025accuracy}. Rafe and Das coded \CompanionCoded{} Texas narratives with a calibrated decision model and audited its probabilities against human judgments~\cite{rafe2026calibrated}. They also screened every Texas narrative from 2017 to 2025 for documented pregnancy, a factor that no coded field records, and corrected the count for misclassification with a blind human adjudication~\cite{rafe2026counting}. Apart from these two, the narrative studies report agreement or prevalence on samples rather than statewide counts, and none of them uses the coded record as a second reader.

Prediction-powered inference gives valid intervals for population quantities from model predictions and a smaller set of trusted measurements. The original method corrects the bias of predictions on unlabeled data with the prediction error on a labeled sample~\cite{angelopoulos2023prediction}, and its power-tuned version asymptotically never does worse than the labeled data alone~\cite{angelopoulos2026ppi}. Cross-prediction trains the predictor on the labeled data by cross-fitting and keeps the guarantees~\cite{zrnic2024cross}, and active inference spends a labeling budget where the model is uncertain~\cite{zrnic2024active}. Design-based supervised learning combines language-model annotations with a probability-sampled gold subset for social science estimands~\cite{egami2023using,chew2026design}. These methods descend from the survey-sampling theory of two-phase designs and calibration~\cite{sarndal1992model,breslow2009improved}. Kluger and Bates extend them to adaptive samples drawn in several waves, with weights that keep each wave unbiased given the earlier ones~\cite{kluger2026estimation}. None of these studies uses two machine readers of different views as nested phases.

Tabular foundation models are transformers pretrained on synthetic tables that predict new rows from labeled context rows in one forward pass~\cite{hollmann2025accurate}. TabICLv2 reported results above tuned baselines on public benchmarks without tuning~\cite{qu2026tabiclv2}, and TabPFN version 3 scaled the context to a million rows while version 3.5 kept that scale for wider tables~\cite{grinsztajn2026tabpfn,jager2026tabpfn}. Kumo Tabular was released with three sizes, class probabilities, and a license for commercial use~\cite{qu2026nvidia}. A living benchmark found these models strong on small and medium tables drawn independently~\cite{erickson2025tabarena}, and the BeyondArena benchmark found them weaker under temporal and grouped shifts~\cite{purucker2026beyond}. In crash research TabPFN has served as a severity classifier on a curated subset~\cite{somvanshi2025tabular}, and a feature-group tabular transformer has predicted collision types~\cite{lares2025feature}. The deep tabular model TabNet predicted pedestrian crash severity in Utah more accurately overall than a generalized ordered probit and a stacking ensemble~\cite{rafe2024comparative}. Severity models have also added officer narratives~\cite{majidi2025predicting} or text generated from the coded fields~\cite{zhen2025crashsage}. These are prediction studies, while the present study uses a tabular model as a calibrated reader inside an inference design. A cascade that routes uncertain items from a decision model to a language model was evaluated for text annotation and suggested as an annotation stage~\cite{ibrahim2026evaluating}, on one view and without an inference target.

Table~\ref{tab:positioning} sets the studies whose methods the present design combines against twelve properties, on which the present study is coded full for ten and partial for two. Among the narrative studies only these two report statewide counts, the inference studies use a single proxy without crash data, and the tabular-model studies serve prediction rather than inference. No study in the table links the two views of the same record under a sampling design, maps confirmed discordance by domain, and validates a re-read list on a probability sample, which are the gaps that the present study addresses.

\begin{table}[t]
\caption{What prior studies did and what this study does.}\label{tab:positioning}
\tablestyle
\setlength{\tabcolsep}{2pt}
\begin{tabular}{P{0.2350}C{0.0638}C{0.0638}C{0.0638}C{0.0638}C{0.0638}C{0.0638}C{0.0638}C{0.0638}C{0.0638}C{0.0638}C{0.0638}C{0.0638}}
\toprule
\headrow \hdtop{} & \multicolumn{3}{c}{\hdtop{Views}} & \multicolumn{2}{c}{\hdtop{Readers}} & \multicolumn{3}{c}{\hdtop{Inference}} & \multicolumn{4}{c}{\hdtop{Outputs}} \\
\headrow \hdsub{Study} & \hdsub{\rotatebox{90}{\shortstack[l]{Narrative\\text read}}} & \hdsub{\rotatebox{90}{\shortstack[l]{Coded\\table read}}} & \hdsub{\rotatebox{90}{\shortstack[l]{Both views\\linked}}} & \hdsub{\rotatebox{90}{\shortstack[l]{In-context\\table model}}} & \hdsub{\rotatebox{90}{\shortstack[l]{Calibration\\audited}}} & \hdsub{\rotatebox{90}{\shortstack[l]{Population\\intervals}}} & \hdsub{\rotatebox{90}{\shortstack[l]{Design\\weights}}} & \hdsub{\rotatebox{90}{\shortstack[l]{Multiwave\\sampling}}} & \hdsub{\rotatebox{90}{\shortstack[l]{Population\\counts}}} & \hdsub{\rotatebox{90}{\shortstack[l]{Discordance\\mapped}}} & \hdsub{\rotatebox{90}{\shortstack[l]{Re-read list\\validated}}} & \hdsub{\rotatebox{90}{\shortstack[l]{Budget\\allocated}}} \\
\midrule
\grouprow\multicolumn{13}{l}{\grouplabel{Crash narratives and crash data}}\\
\citet{bharati2026benchmarking} & \symfull & \symnone & \symhalf & \symnone & \symnone & \symnone & \symnone & \symnone & \symnone & \symhalf & \symnone & \symnone \\
\citet{majidi2025predicting} & \symfull & \symfull & \symfull & \symnone & \symnone & \symnone & \symnone & \symnone & \symnone & \symnone & \symnone & \symnone \\
\citet{nhtsa2025distracted} & \symnone & \symnone & \symnone & \symnone & \symnone & \symhalf & \symhalf & \symnone & \symfull & \symnone & \symnone & \symnone \\
\citet{rafe2026calibrated} & \symfull & \symnone & \symhalf & \symnone & \symfull & \symhalf & \symhalf & \symhalf & \symfull & \symhalf & \symhalf & \symfull \\
\citet{rafe2026counting} & \symfull & \symhalf & \symhalf & \symnone & \symhalf & \symhalf & \symfull & \symfull & \symfull & \symnone & \symfull & \symhalf \\
\citet{rolison2018what} & \symnone & \symnone & \symnone & \symnone & \symnone & \symnone & \symnone & \symnone & \symhalf & \symnone & \symnone & \symnone \\
\citet{saroj2026adapting} & \symnone & \symfull & \symnone & \symnone & \symnone & \symnone & \symnone & \symnone & \symnone & \symnone & \symnone & \symnone \\
\citet{zhen2025crashsage} & \symnone & \symfull & \symnone & \symnone & \symnone & \symnone & \symnone & \symnone & \symnone & \symnone & \symnone & \symnone \\
\grouprow\multicolumn{13}{l}{\grouplabel{Inference with model predictions}}\\
\citet{angelopoulos2023prediction} & \symnone & \symhalf & \symnone & \symnone & \symnone & \symfull & \symhalf & \symnone & \symfull & \symnone & \symnone & \symnone \\
\citet{angelopoulos2026ppi} & \symnone & \symhalf & \symnone & \symnone & \symnone & \symfull & \symnone & \symnone & \symhalf & \symnone & \symnone & \symnone \\
\citet{chew2026design} & \symhalf & \symhalf & \symhalf & \symnone & \symnone & \symfull & \symfull & \symfull & \symhalf & \symnone & \symhalf & \symhalf \\
\citet{egami2023using} & \symfull & \symhalf & \symhalf & \symnone & \symnone & \symfull & \symfull & \symhalf & \symhalf & \symnone & \symnone & \symnone \\
\citet{kluger2026estimation} & \symnone & \symhalf & \symnone & \symnone & \symnone & \symfull & \symfull & \symfull & \symhalf & \symnone & \symnone & \symhalf \\
\citet{zrnic2024cross} & \symnone & \symhalf & \symnone & \symnone & \symnone & \symfull & \symnone & \symnone & \symhalf & \symnone & \symnone & \symnone \\
\grouprow\multicolumn{13}{l}{\grouplabel{Tabular foundation models}}\\
\citet{hu2026noise} & \symnone & \symfull & \symnone & \symfull & \symnone & \symnone & \symnone & \symnone & \symnone & \symnone & \symnone & \symnone \\
\citet{purucker2026beyond} & \symhalf & \symfull & \symhalf & \symfull & \symhalf & \symnone & \symnone & \symnone & \symnone & \symnone & \symnone & \symnone \\
\citet{rafe2024comparative} & \symnone & \symfull & \symnone & \symnone & \symnone & \symnone & \symnone & \symnone & \symnone & \symnone & \symnone & \symnone \\
\citet{shi2026multi} & \symnone & \symfull & \symnone & \symfull & \symhalf & \symnone & \symnone & \symhalf & \symnone & \symnone & \symnone & \symhalf \\
\citet{somvanshi2025applying} & \symnone & \symfull & \symnone & \symfull & \symnone & \symnone & \symnone & \symnone & \symnone & \symnone & \symnone & \symnone \\
\citet{somvanshi2025tabular} & \symnone & \symfull & \symnone & \symfull & \symnone & \symnone & \symnone & \symnone & \symnone & \symnone & \symnone & \symnone \\
\grouprow\multicolumn{13}{l}{\grouplabel{Decision-model cascades}}\\
\citet{ibrahim2026evaluating} & \symfull & \symnone & \symnone & \symnone & \symfull & \symnone & \symnone & \symnone & \symnone & \symnone & \symhalf & \symfull \\
\midrule
Prior studies, full\,/\,partial & 6\,/\,2 & 9\,/\,7 & 1\,/\,6 & 5\,/\,0 & 2\,/\,3 & 6\,/\,3 & 4\,/\,3 & 3\,/\,3 & 4\,/\,6 & 0\,/\,2 & 1\,/\,3 & 2\,/\,4 \\
\midrule
\textbf{This study} & \symfull & \symfull & \symfull & \symfull & \symfull & \symhalf & \symhalf & \symfull & \symfull & \symfull & \symfull & \symfull \\
\bottomrule
\end{tabular}
\tablenote{\symfull{} full, \symhalf{} partial, \symnone{} absent, each coded from the study's own full text and, for this study, by the same rules. The coverage row counts the prior studies that meet each criterion fully and partly. The supplementary evidence file gives, for every cell, a verbatim supporting quotation and its location or, for an absent cell, the search terms that found nothing.}
\end{table}

\section{Data and design}\label{sec:data}

The data are the Texas Crash Records Information System extract for 2017 to 2025, which holds \NAll{} crashes in its person-level file. Each crash carries a coded record built from the crash, unit, and person tables, and most crashes carry the investigating officer's narrative. We define the narrative population $U$ as the geolocated crashes whose narrative has more than forty characters. It contains \NU{} crashes, or \ShareGeoU{} percent of all crashes. Every estimate in this paper refers to $U$ or to a domain of it, and the crashes outside $U$ are reported separately so that no total silently mixes the two.

\begin{figure}[t]
\centering
\includegraphics[width=\textwidth]{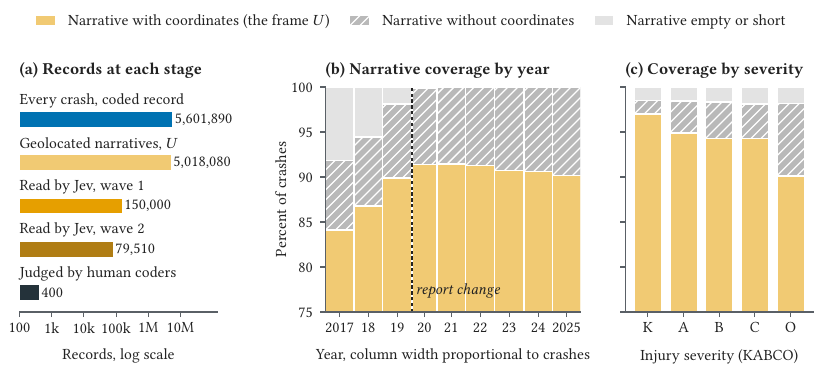}
\caption{The population and its coverage. Panel (a) gives the size of each phase on a log scale. Panel (b) splits each year's crashes into the narrative population $U$, crashes with a narrative but no coordinates, and crashes whose narrative is empty or short, with column width proportional to the crashes of that year. Panel (c) gives the same split by severity on the KABCO scale, from fatal (K) to no injury (O). Both share axes start at \CovAxisFloor{} percent.}\label{fig:coverage}
\end{figure}

The crashes outside $U$ fall into two groups whose reasons differ, and Figure~\ref{fig:coverage} shows both by year and by severity. The larger group holds \NNonGeo{} crashes whose narrative exists but which lack coordinates on every record. The narrative export kept only geolocated crashes, so these narratives were never read. The smaller group holds \NEmptyNarr{} crashes whose narrative is empty or at most forty characters long, and these cluster in 2017 to 2019, before the crash report change of 2020. A further \NLongWithCoords{} crashes have a long narrative and coordinates but were absent from the export. Crashes outside $U$ are less severe on average, with a fatal share of \FatalShareOutU{} percent against \FatalShareU{} percent inside $U$, yet they include \NFatalOutU{} fatal crashes.

The coded record that the table reader sees has \NFeatures{} crash, unit, and person features, of which \NFeaturesCat{} are categorical (Appendix~\ref{app:features}). Missing cells stay missing, no date finer than the year enters, and identifiers, coordinates, and narrative text never enter.

Table~\ref{tab:phases} lists the phases of the design. Phase 0 is the table reader's output for every crash. Phase 1 is the narrative reader's output on two probability samples of $U$, a simple random sample of \nOne{} narratives and a second wave of \nTwoDraws{} narratives drawn with recorded probabilities. Phase 2 is the human tier, \nGoldJudg{} usable judgments on \nGoldNarr{} narratives drawn from the random sample by bins of the narrative reader's probability. A judgment is the consensus label of two or three coders, who did not see the model output, for one narrative and one factor. The check is a second human tier drawn with recorded probabilities. Narratives read in enriched samples, whose probabilities are conditional on their draw pool, stay out of wave 1, and wave 2 reuses their reads when it draws them. The wave-1 reads and the phase-2 human tier were also analyzed, for different questions, by Rafe and Das~\cite{rafe2026calibrated}.

\begin{table}[t]
\caption{The phases of the design and what is known at each.}\label{tab:phases}
\tablestyle
\begin{tabular}{P{0.15}P{0.17}P{0.14}P{0.13}P{0.16}P{0.25}}
\toprule
\headrow \hd{Phase} & \hd{Units} & \hd{Size} & \hd{Reader} & \hd{Measurement} & \hd{Design information} \\
\midrule
Phase 0 & Every crash & 5,601,890 & Kumo Tabular & $\hat f_i$ for every factor & Complete enumeration \\
Phase 1, wave 1 & Random stratum of $U$ & 150,000 & Jev & $p_i$, recalibrated to $\tilde q_i$ & $\pi^{(1)} = n_1/N$, simple random sample \\
Phase 1, wave 2 & $U$ outside wave 1 & 79,510 & Jev & $p_i$, recalibrated to $\tilde q_i$ & Recorded $\pi^{(2)}_i$, Poisson draw on the discordance score \\
Phase 2 & Human-judged narratives & 400 (2,309 judgments) & Two or three coders & $y_i$ & Calibration weights, model-dependent \\
Check & Wave-1 narratives outside phase 2 & 600 (427 read by a specialist) & Three coders & $y_i$ & Recorded inclusion probabilities, Poisson draw, design-based \\
\bottomrule
\end{tabular}
\tablenote{$U$ is the set of geolocated crashes with a narrative of more than forty characters. Wave 2 reused the existing reads of the narratives it drew from earlier enriched samples.}
\end{table}

\section{Methods}\label{sec:methods}

\begin{figure}[t]
\centering
\includegraphics[width=\textwidth]{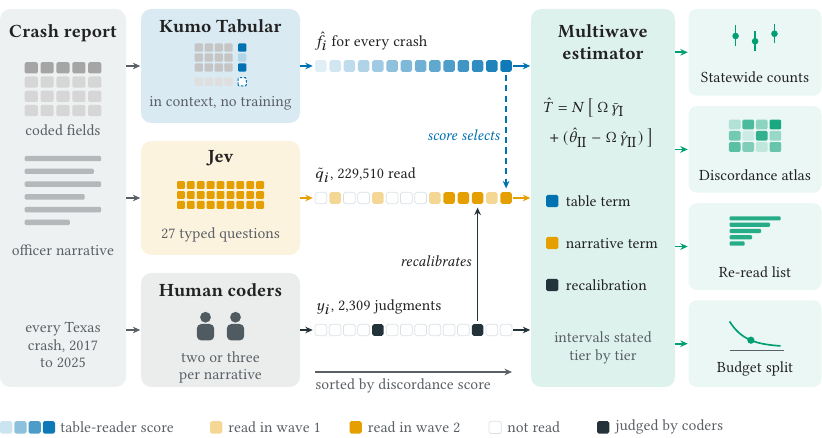}
\caption{The two-view design. Each column of cells is one crash, sorted by the table reader's discordance score, and the cells show which crashes each reader and the human coders touch. The cells are schematic, and the counts are design facts.}\label{fig:design}
\end{figure}

The methods combine two decision models and a human tier through a sampling design. The table reader returns a probability for every crash from the coded record, the narrative reader returns a probability for every narrative it reads, and the human tier recalibrates the narrative reader. The estimator treats the narrative reading as a phase-1 measurement, the table reading as a proxy known for the whole population, and the human judgments as the reference for what a narrative states. Figure~\ref{fig:design} shows how the two readers, the human tier, and the estimator fit together, and Table~\ref{tab:notation} in Appendix~\ref{app:estimator} collects the notation.

\subsection{Readers, estimator, and human check}\label{sec:readers}\label{sec:estimator}

The narrative reader is Jev, pinned to version jev-1.13.0, with a typed schema of \SchemaQuestions{} questions and its gating rule~\cite{rafe2026calibrated}. The table reader is Kumo Tabular~\cite{qu2026nvidia}, pinned to weight revision v1.0.0 and run in its small size with a context of \CtxRows{} labeled rows. It receives the coded record as numerical and categorical columns, keeps missing cells as missing, and returns the probability that the narrative reader flags the factor.

The narrative reader's probabilities are recalibrated against the human tier before they enter the estimator. For factor $v$ the map is
\begin{equation}\label{eq:recal}
\operatorname{logit} g_v(p, c) = a_{k(v)} + b_{k(v)} \operatorname{logit}(p) + \eta\, c\, \mathbf{1}[v \text{ has a coded field}],
\end{equation}
where $k(v)$ is the factor itself when it has at least twenty human positives and a pooled group otherwise, and $\eta$ is one coded-field shift shared by the nine factors with a coded counterpart. The shift is shared because no factor reaches twenty human positives on both sides of its coded field. We fitted Eq.~\eqref{eq:recal} by weighted logistic regression with the calibration weights of the human tier. Against the identity, a plain Platt map~\cite{platt2000probabilities}, and an isotonic map, a held-out rule admitted only maps unbiased in the mean and then kept the simplest within one standard error of the best Brier score~\cite{brier1950verification} (Appendix~\ref{app:recal}).

The table reader is cross-fitted so that no crash informs its own prediction~\cite{zrnic2024cross}. The wave-1 sample is split into \Kfolds{} folds, the fold-$k$ model takes its context from the other folds only, a wave-1 crash receives the prediction of the model whose context excludes it, and every other crash receives the average of the fold models. Each fold model is fitted on the narrative reader's flag $\mathbf{1}[p_{vi} \ge 0.5]$, because recalibration moves the rare factors below one half and would leave too few positives in a context, and the estimator stays valid for any reader output.

The estimand $T_{v\delta} = \sum_{i \in U} \delta_i\, \tilde q_{vi}$ is the number of crashes in a domain whose narrative states $v$, read through the recalibrated narrative reader. Wave 1 is a simple random sample of $n_1 = \nOne{}$ narratives from $U$, and wave 2 is a Poisson sample from the remaining narratives drawn on the table reader's discordance score (Section~\ref{sec:wave2}). In the multiwave construction of Kluger and Bates~\cite{kluger2026estimation}, each wave carries its own inverse probability weight, and the two weights are combined with fixed shares as follows.
\begin{equation}\label{eq:weights}
W_i = c_1 \frac{I^{(1)}_i}{\pi^{(1)}_i} + c_2 \frac{1 - I^{(1)}_i}{1 - \pi^{(1)}_i}\, \frac{I^{(2)}_i}{\pi^{(2)}_i}, \qquad c_1 + c_2 = 1.
\end{equation}
The share $c_1 = \cOne{}$ was set in advance to the share of wave 1 in the expected number of measured narratives. Each wave term has expectation one for every unit, so the combination neither double counts nor drops a unit.

The multiwave predict-then-debias estimator anchors on the table reader over the whole population and corrects its bias with the weighted measured narratives.
\begin{equation}\label{eq:mpd}
\hat T_{v\delta} = N \Big[\, \Omega\, \bar\gamma_{\mathrm{I}} + \big(\hat\theta_{\mathrm{II}} - \Omega\, \hat\gamma_{\mathrm{II}}\big) \Big], \quad
\hat\theta_{\mathrm{II}} = \frac{\sum_i W_i \delta_i \tilde q_{vi}}{\sum_i W_i}, \quad
\hat\gamma_{\mathrm{II}} = \frac{\sum_i W_i \delta_i \hat f_{vi}}{\sum_i W_i}, \quad
\bar\gamma_{\mathrm{I}} = \frac{1}{N}\sum_{i \in U} \delta_i \hat f_{vi}.
\end{equation}
For $\Omega = 0$ the estimator is the weighted mean of the measured narratives, and the two terms in $\Omega$ share their limit whatever the table reader, so the estimator is consistent for any reader. We choose $\Omega$ to minimize the variance estimator of Kluger and Bates and restrict it to the unit interval as in PPI++~\cite{angelopoulos2026ppi}. Appendix~\ref{app:estimator} gives the variance, and every wave-2 probability lies between \PiTwoMin{} and \PiTwoMax{}, which meets the overlap condition of the construction.

The human tier enters through $g_v$, so the phase-1 part is design-based and the human tier is model-dependent. The uncertainty of $g_v$, map selection included, is inside every interval, through \BRecal{} narrative cluster resamples of the human judgments that repeat the pooling rule, the selection of the map, and its fit (Appendix~\ref{app:recal}). The sensitivity term, the weighted mean of the human label minus the recalibrated probability over the human tier, scaled to the domain, is reported beside each estimate.

A second human tier checks every estimate by design. It is a Poisson sample of the \RtwoFrame{} wave-1 narratives that the phase-2 tier had not judged, drawn with recorded inclusion probabilities after the estimates were fixed, and \RtwoCoded{} of its narratives were coded after an identifier screen (Appendix~\ref{app:round2}). Three coders who saw no model output took part, two specialists in crash data and safety research and a careful reader without crash background. The specialists agreed on \AgreeSpec{} percent of their shared judgments (Cohen's $\kappa = \KappaSpec{}$), while the third coder answered yes \NonspecYesShare{} percent as often. The reference label is therefore the specialist's, as the agreement of coders who agreed on \RoneAgree{} percent of pairs is the reference of the phase-2 tier. The check adds to each estimate the correction $\hat C_{v} = \sum_j w_j \delta_j (y_{vj} - \tilde q_{vj})$ over the judged factors that the narrative reader gave a probability of at least \ActiveThreshold{}, where $w_j$ is the inverse of the product of the inclusion probabilities. Because the second tier never entered the recalibration, $\hat T_{v\delta} + \hat C_v$ targets the human criterion by design in that region. Below it lies at most \LowpOtherMax{} percent of any estimate, except \LowpWrongWay{} percent for wrong-way driving and \LowpDefect{} percent for vehicle defects.

Two comparison estimators show what each component adds, the wave-1-only estimator, which sets $W_i = I^{(1)}_i / \pi^{(1)}_i$, and the sequential estimator of Appendix~\ref{app:wave2}, and a simulation checked the coverage of all three before any real number was computed. For the \NEmptyNarr{} crashes whose narrative is empty or at most forty characters and the \NNonGeo{} crashes without coordinates, the sum of $\hat f_{vi}$ is reported apart from every estimate as a model-based extrapolation.

\subsection{Second wave, outputs, and table-reader evaluation}\label{sec:wave2}\label{sec:atlas}\label{sec:allocation}\label{sec:eval}

The second wave reads narratives where the coded record suggests that the narrative states a factor that the field does not. Every narrative of $U$ outside wave 1 received the score $s_i = \sum_{v} \hat f^{\,d}_{vi} / \bar f^{\,d}_{v}$, where $\hat f^{\,d}_{vi}$ is the table reader's probability of confirmed discordance for factor $v$ and $\bar f^{\,d}_{v}$ is its mean over the frame, so a rare factor weighs as much as a common one. Each narrative was drawn independently with probability
\begin{equation}\label{eq:pi2}
\pi^{(2)}_i = \min\Big\{1 - \big(1 - \min(1, \kappa s_i)\big)\big(1 - c_0\big),\ \PiCap\Big\},
\end{equation}
the probability of selection by the score-driven draw or by a uniform control draw with probability $c_0$. The values of $\kappa$ and $c_0$ were solved for \nTwoExpScore{} and \nTwoExpControl{} expected units, and the draw, made once from a fixed seed, produced \nTwoDraws{} narratives, of which \nTwoReuse{} reused an existing read and \nTwoNew{} were read with the same schema and model.

Confirmed discordance is the event that a human would say the narrative states a factor that the coded field does not record. Its phase-1 measurement is $\tilde q_{vi}(1 - c_{vi})$, and its table-reader proxy is a separate cross-fitted Kumo Tabular model of the discordance flag $\mathbf{1}[p_{vi} \ge 0.5]\,(1 - c_{vi})$. Discordance is not coding error, because a field may lag the narrative for procedural reasons, such as a test result that alcohol requires. We therefore use narrative surplus for the information that only the narrative carries, and we estimate discordance for the nine factors with a coded field.

The atlas reports domain totals and rates of confirmed discordance by year, road system, population group, setting, severity, road class, and county, and each total is Eq.~\eqref{eq:mpd} with the domain indicator. A contrast between two domains is the ratio of their rates, with an interval that recomputes the ratio within each recalibration refit, so that the recalibration uncertainty shared by the two domains cancels. A county is pooled into one cell when its crash count times the statewide rate falls below ten.

The re-read list ranks the unread narratives by the table reader's discordance probability. Its precision at the top $k$ percent is the sequential estimate of confirmed discordance among the listed crashes divided by the list size, and its lift is that precision over the rate among all unread narratives. Because the second wave has recorded probabilities that are positive everywhere, these are design-based estimates, with confirmation through the human-recalibrated reader.

The two variance components of each estimate fall with different purchases. The design variance falls with the number of narratives read, and the recalibration variance with the number of human-coded narratives.
\begin{equation}\label{eq:alloc}
\operatorname{Var}(m_1, m_2) = \frac{A_v}{m_1}\Big(1 - \frac{m_1}{N}\Big) + \frac{B_v}{m_2}, \qquad m_1 \propto \sqrt{A_v / u_1}, \quad m_2 \propto \sqrt{B_v / u_2},
\end{equation}
Here $m_1$ and $m_2$ are the numbers of narratives read and human-coded, $A_v$ and $B_v$ the measured design and recalibration variances multiplied by their current sizes, and $u_1$ and $u_2$ the price of one read and the planning cost of one human-coded narrative. The proportionality is the Neyman allocation~\cite{neyman1934two,neyman1938contribution} applied to the two phases under the budget $u_1 m_1 + u_2 m_2$.

The table-reader evaluation compares in-context foundation models with a trained booster on the same contexts, test rows, and labels. The test set is one fold of wave 1, and contexts of increasing size are nested prefixes of the remaining folds. The arms are Kumo Tabular in three sizes, TabICLv2~\cite{qu2026tabiclv2}, TabPFN 3 and TabPFN 3.5~\cite{grinsztajn2026tabpfn,jager2026tabpfn} under their evaluation license, and CatBoost~\cite{prokhorenkova2018catboost} trained on each context and on the whole pool. We measure discrimination, calibration with standard metrics~\cite{rafe2026calibrated}, throughput, and memory on one A100 GPU, with CatBoost on the CPU of the same machine. The production reader is also compared over all five folds with a booster tuned per target, and both are scored against the human labels.

\section{Results}\label{sec:results}

\subsection{What the narrative carries and how many crashes it documents}\label{sec:res-plane}\label{sec:res-estimates}

\begin{figure}[t]
\centering
\includegraphics[width=\textwidth]{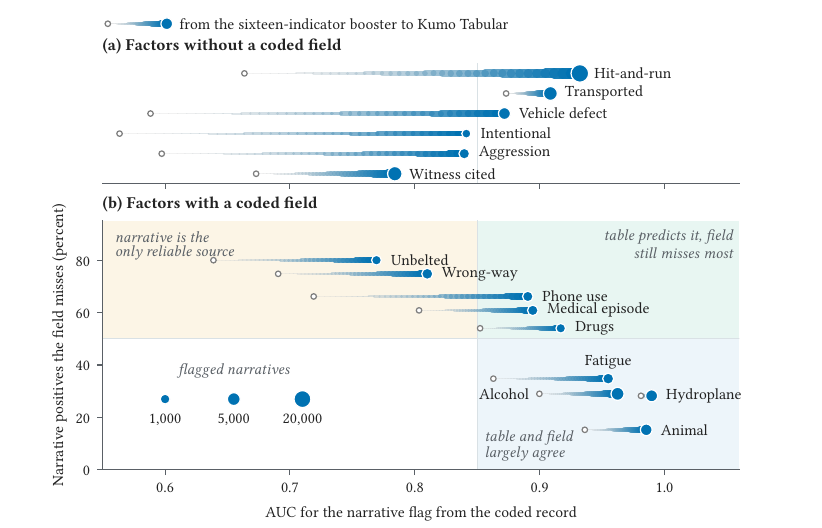}
\caption{Predictability of the narrative flag from the coded record against the share of narrative positives that the coded field misses. Panel (a) gives the factors without a coded field and panel (b) those with one. Each comet runs from the AUC of a booster on sixteen coded factor indicators (tail) to the AUC of Kumo Tabular on the full coded record (head), and the head grows with the number of flagged wave-1 narratives, as the key in panel (b) shows. The lines at an AUC of \AucLine{} and at half of the narrative positives divide the plane into the labeled regions.}\label{fig:plane}
\end{figure}

The two views are complementary in a way that can be measured factor by factor, which answers RQ1. Figure~\ref{fig:plane} places each factor by how well the coded record predicts the narrative flag and by the share of narrative positives that the coded field misses, and Table~\ref{tab:predictability} gives the values. For the nine factors with a coded field the points fall from the upper left to the lower right. Animal, hydroplane, alcohol, and fatigue are predicted with an AUC of \KumoAucFatigue{} or more, and their fields miss between \NarrOnlyAnimal{} and \NarrOnlyFatigue{} percent of what the narrative states. Unbelted occupants and wrong-way driving sit at the opposite corner, with AUCs of \KumoAucUnbelted{} and \KumoAucWrongWay{} and fields that miss \NarrOnlyUnbelted{} and \NarrOnlyWrongWay{} percent.

The full coded record predicts much more than sixteen coded factor indicators, with the largest gains among the coded factors for phone use, from \AucSixteenPhone{} to \KumoAucPhone{}, and for unbelted occupants, from \AucSixteenUnbelted{} to \KumoAucUnbelted{}. The record also carries factors without a coded field indirectly, since hit-and-run reaches an AUC of \KumoAucHitRun{} and transport to a hospital \KumoAucTransported{}, while witness citations remain the least predictable of these factors, at \KumoAucWitness{}. Where the table predicts the narrative well, the table reader is an efficient control variate, and where the narrative surplus is large, the atlas and the re-read list matter most.

\begin{table}[t]
\caption{Predictability of the narrative reading from the coded record, and the narrative surplus.}\label{tab:predictability}
\tablestyle
\begin{tabular}{P{0.17}R{0.08}R{0.09}R{0.08}R{0.08}R{0.20}R{0.15}R{0.15}}
\toprule
\headrow \hd{} & \hd{} & \multicolumn{3}{c}{\hdtop{AUC for the narrative flag}} & \multicolumn{3}{c}{\hdtop{Narrative surplus}} \\
\headrow \hdsub{Factor} & \hdsub{Flags} & \hdsub{Sixteen} & \hdsub{Full} & \hdsub{Kumo} & \hdsub{Field misses} & \hdsub{Confirmed misses} & \hdsub{AUC, field false} \\
\midrule
\grouprow\multicolumn{8}{l}{\grouplabel{Factors with a coded field}}\\
Unbelted$^{\dagger}$ & 933 & 0.64 & 0.73 & \best{0.77} & \cellbar[discver]{0.80}{80\%} & 37\% & 0.69 \\
Wrong-way & 2,199 & 0.69 & 0.80 & \best{0.81} & \cellbar[discver]{0.75}{75\%} & 44\% & 0.72 \\
Phone use$^{\dagger}$ & 1,562 & 0.72 & 0.88 & \best{0.89} & \cellbar[discver]{0.66}{66\%} & 65\% & 0.81 \\
Medical episode & 1,608 & 0.80 & \best{0.89} & \best{0.89} & \cellbar[discver]{0.61}{61\%} & 47\% & 0.78 \\
Drugs$^{\dagger}$ & 850 & 0.85 & 0.86 & \best{0.92} & \cellbar[discver]{0.54}{54\%} & 38\% & 0.64 \\
Fatigue & 2,101 & 0.86 & \best{0.95} & \best{0.95} & \cellbar[discver]{0.35}{35\%} & 33\% & 0.77 \\
Alcohol & 5,470 & 0.90 & \best{0.96} & \best{0.96} & \cellbar[discver]{0.29}{29\%} & 22\% & 0.86 \\
Hydroplane & 3,547 & 0.98 & \best{0.99} & \best{0.99} & \cellbar[discver]{0.28}{28\%} & 26\% & 0.96 \\
Animal & 3,028 & 0.94 & 0.98 & \best{0.99} & \cellbar[discver]{0.15}{15\%} & \cellnote{none} & 0.78 \\
\grouprow\multicolumn{8}{l}{\grouplabel{Factors without a coded field}}\\
Hit-and-run & 23,859 & 0.66 & \best{0.93} & \best{0.93} & \cellnote{none} & \cellnote{none} & \cellnote{none} \\
Transported & 7,509 & 0.87 & \best{0.91} & \best{0.91} & \cellnote{none} & \cellnote{none} & \cellnote{none} \\
Vehicle defect & 3,351 & 0.59 & 0.86 & \best{0.87} & \cellnote{none} & \cellnote{none} & \cellnote{none} \\
Intentional$^{\dagger}$ & 562 & 0.56 & 0.76 & \best{0.84} & \cellnote{none} & \cellnote{none} & \cellnote{none} \\
Aggression & 1,688 & 0.60 & 0.82 & \best{0.84} & \cellnote{none} & \cellnote{none} & \cellnote{none} \\
Witness cited & 9,279 & 0.67 & \best{0.78} & \best{0.78} & \cellnote{none} & \cellnote{none} & \cellnote{none} \\
\bottomrule
\end{tabular}
\tablenote{Flags are Jev flags at $p \ge 0.5$ in the \nOne{} wave-1 narratives. Sixteen and Full are out-of-fold gradient-boosting AUCs from a booster on the sixteen coded indicators and on the full coded record, and Kumo is the out-of-fold Kumo Tabular AUC used by the estimator. Bold marks the highest AUC in each row at printed precision, with ties bolded. Field misses is the share of flagged narratives whose coded field is false, drawn as a bar whose full track is 100 percent. Confirmed misses weights each side of the coded field by its human confirmation rate in Table~\ref{tab:recal}, and is none where the human tier holds no flag with the field false. The last column is the booster AUC for that event among crashes whose field is false. $^{\dagger}$Pooled recalibration map.}
\end{table}

The narrative documents more injury crashes than the coded field for five factors, which answers RQ2. Table~\ref{tab:estimates} gives the number of injury-or-fatal crashes from 2017 to 2025 whose narrative states each factor, and for hydroplaning, medical episodes, fatigue, animals, and phone use the whole interval lies above the coded count. Phone use is the clearest case, with \EstPhone{} narrative-stated crashes against \CodedPhone{} coded ones, and medical episodes follow with \EstMedical{} against \CodedMedical{}. Transport to a hospital, hit-and-run, and witness citations are the most frequent factors without a coded field, with \EstTransported{}, \EstHitRun{}, and \EstWitness{} crashes. The coded count is close to or above the narrative count for alcohol, at \CodedAlcohol{} against \EstAlcohol{}, and far above it for unbelted occupants, at \CodedUnbelted{} against \EstUnbelted{}, a comparison of a driver-level field with an occupant-level narrative statement rather than an error rate of either view.

\begin{table}[t]
\caption{Injury-or-fatal crashes in $U$ whose narrative states each factor, 2017 to 2025, with the design-based human check.}\label{tab:estimates}
\tablestyle
\begin{tabular}{P{0.19}R{0.12}R{0.19}R{0.12}R{0.12}R{0.13}R{0.13}}
\toprule
\headrow \hd{} & \multicolumn{3}{c}{\hdtop{Narrative states the factor}} & \hd{} & \multicolumn{2}{c}{\hdtop{Second human tier}} \\
\headrow \hdsub{Factor} & \hdsub{Estimate} & \hdsub{95 percent interval} & \hdsub{Sensitivity} & \hdsub{Coded field} & \hdsub{Human check} & \hdsub{95 percent interval} \\
\midrule
Transported & 224,996 & 208,813 to 234,969 & -1,062 & \cellnote{none} & 241,450 & 216,770 to 266,130 \\
Hit-and-run & 143,701 & 135,642 to 164,848 & -896 & \cellnote{none} & 102,984 & 4,368 to 201,600 \\
Witness cited & 135,504 & 115,809 to 152,568 & -252 & \cellnote{none} & 157,977 & 132,589 to 183,364 \\
Alcohol & 65,971 & 60,078 to 73,566 & -699 & 73,809 & 69,802 & 65,188 to 74,416 \\
Vehicle defect & 59,396 & 35,923 to 60,699 & -2,087 & \cellnote{none} & 61,595 & 45,927 to 77,262 \\
Hydroplane & \best{32,996} & 30,589 to 35,657 & 159 & 27,431 & 31,941 & 29,883 to 33,999 \\
Medical episode & \best{28,345} & 21,793 to 31,120 & -35 & 16,193 & 28,825 & 25,834 to 31,816 \\
Fatigue & \best{24,880} & 22,784 to 25,907 & -508 & 21,735 & 26,568 & 22,466 to 30,669 \\
Animal & \best{20,195} & 18,589 to 21,010 & 382 & 15,691 & 19,662 & 18,620 to 20,704 \\
Wrong-way & 18,294 & 12,639 to 25,820 & 538 & 14,338 & 20,233 & 12,142 to 28,324 \\
Phone use$^{\dagger}$ & \best{15,074} & 13,037 to 18,186 & 468 & 7,340 & 16,211 & 13,050 to 19,371 \\
Drugs$^{\dagger}$ & 12,664 & 9,578 to 13,595 & -21 & 14,883 & 14,105 & 10,369 to 17,840 \\
Aggression & 10,795 & 8,879 to 12,798 & 175 & \cellnote{none} & 13,958 & 8,807 to 19,108 \\
Unbelted$^{\dagger}$ & 9,617 & 7,661 to 11,355 & -767 & 57,657 & 11,354 & 6,781 to 15,926 \\
Intentional$^{\dagger}$ & 1,870 & 1,254 to 3,030 & 573 & \cellnote{none} & 3,625 & 865 to 6,385 \\
\bottomrule
\end{tabular}
\tablenote{Two-wave multiwave estimates over the injury-or-fatal crashes of $U$. Each interval bound adds the design half-width and the spread of the recalibration refits, map selection included. Sensitivity is the weighted mean of the human label minus the recalibrated probability over the human tier, scaled to the domain, a check on what recalibration leaves. The coded-field count is the number of injury-or-fatal crashes in $U$ whose coded field records the factor. Bold marks an estimate whose whole interval lies above the coded-field count. The human check adds to the estimate the design-based correction from the second human tier, a probability sample whose specialist labels never entered the recalibration, so it does not rest on the recalibration map where the narrative reader gives the factor a probability of at least 0.05. $^{\dagger}$Pooled recalibration map.}
\end{table}

\begin{figure}[t]
\centering
\includegraphics[width=\textwidth]{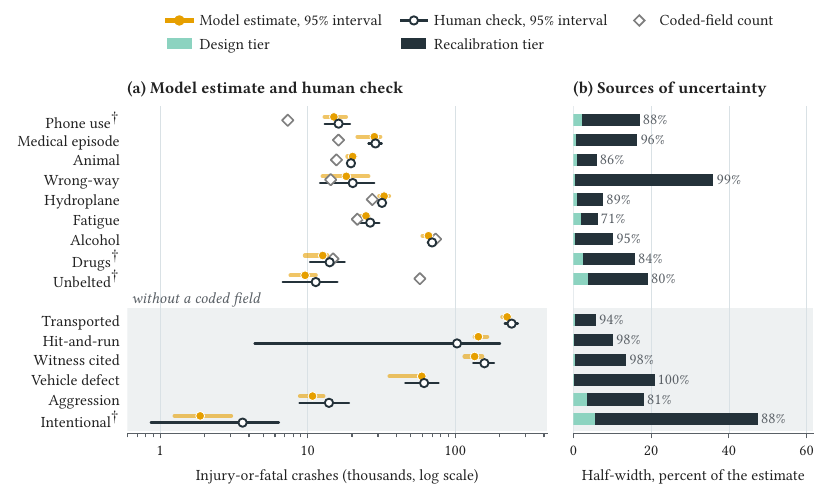}
\caption{Model estimates, their design-based human check, and the sources of uncertainty for injury-or-fatal crashes, 2017 to 2025. Panel (a) gives, on a log scale, the model estimate with its 95 percent interval (amber), the human check of the second human tier with its interval (open circle), and the coded-field count (diamond) where one exists. Panel (b) splits the half-width of each model estimate by variance share into the design tier and the recalibration tier, with the recalibration share printed. $^{\dagger}$Pooled recalibration map.}\label{fig:anatomy}
\end{figure}

The second human tier agrees with these counts. For all \RtwoConsistent{} factors the design-based correction lies within its 95 percent margin of zero, so on every row of Figure~\ref{fig:anatomy}(a) the interval of the human check contains the model estimate. Table~\ref{tab:estimates} gives the values in its last two columns, and the check is precise to within a quarter of the estimate for \RtwoNInformative{} factors. The \RtwoNAbove{} factors whose whole interval lies above the coded count stay above it under the check, with \RtwoPhone{} phone-use crashes (95 percent interval \RtwoPhoneLo{} to \RtwoPhoneHi{}) and \RtwoMedical{} medical-episode crashes (\RtwoMedicalLo{} to \RtwoMedicalHi{}). The specialist reference matters, since the third coder's labels alone would have put alcohol at \RtwoNonspecAlcohol{} crashes. Under the agreement rule of the phase-2 tier, which here drops every pair in which the third coder missed a statement, \RtwoConsConsistent{} of the corrections stay within their margins. The agreement is weakest evidence for vehicle defects, \LowpDefect{} percent of whose estimate lies below the probability threshold of the check, and for hit-and-run and intentional crashes, whose check intervals are widest.

Panel (b) of Figure~\ref{fig:anatomy} shows where the uncertainty of the model estimates comes from, since the recalibration tier carries most of the variance, \RecalShareAlcohol{} percent for alcohol and \RecalShareWrongWay{} percent for wrong-way driving. The table term removes \TableGainMedian{} percent of the design half-width for the median factor and up to \TableGainMax{} percent for \TableGainMaxFactor{}, and the second wave narrows the Wald half-width relative to wave 1 alone, from \OneHwPhone{} to \TwoHwPhone{} percent of the estimate for phone use. With the recalibration tier dominant, however, the total half-width narrows by \TableGainTotalMedian{} percent for the median factor.

\begin{figure}[t]
\centering
\includegraphics[width=\textwidth]{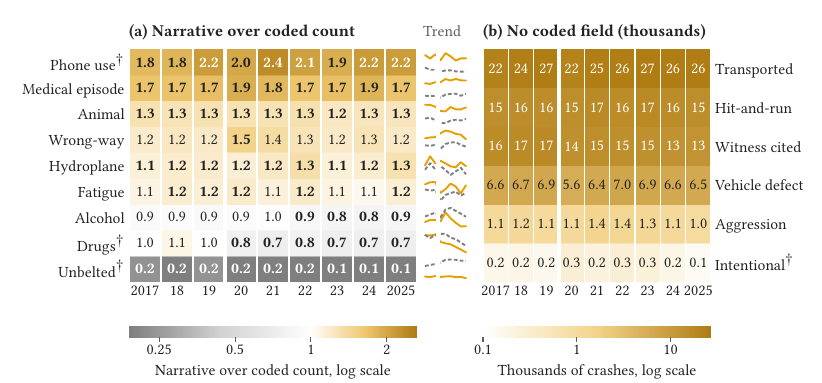}
\caption{Injury-or-fatal crashes per year whose narrative states each factor. Panel (a) gives the narrative estimate over the coded-field count, in bold where the whole 95 percent interval lies on one side of parity, and the trend column draws the two annual series. Panel (b) gives the estimates in thousands for the factors without a coded field. The white seam marks the 2020 change of the crash report. $^{\dagger}$Pooled recalibration map.}\label{fig:yearly}
\end{figure}

The annual series in Figure~\ref{fig:yearly} are stable for most factors. The narrative estimate exceeds the coded count in every year for animals, fatigue, hydroplaning, medical episodes, phone use, and wrong-way driving, and the narrative count of drugs drops in 2020, the year the crash report changed, while the coded count holds through 2021.

\subsection{Discordance, reading priorities, and the table reader}\label{sec:res-atlas}\label{sec:res-budget}\label{sec:res-reader}

Confirmed discordance concentrates where the evidence behind a factor depends on the setting of the crash, which answers RQ3. Table~\ref{tab:atlas} gives the statewide totals and rates for the nine factors with a coded field. Alcohol, hydroplaning, and wrong-way driving have the highest rates, between \AtlasRateWrongWay{} and \AtlasRateAlcohol{} per 1,000 crashes with overlapping intervals, and unbelted occupants the lowest, \AtlasRateUnbelted{}. As a share of the crashes in $U$ whose narrative states the factor, estimated with the recalibrated reader rather than from the flags of Table~\ref{tab:predictability}, confirmed discordance ranges from \CondShareAnimal{} percent for animals to \CondShareWrongWay{} percent for wrong-way driving.

\begin{table}[t]
\caption{Confirmed discordance between narrative and coded field, Texas 2017 to 2025.}\label{tab:atlas}
\tablestyle
\begin{tabular}{P{0.16}R{0.09}R{0.15}R{0.14}R{0.075}R{0.075}R{0.075}R{0.075}R{0.075}R{0.085}}
\toprule
\headrow \hd{} & \multicolumn{3}{c}{\hdtop{Statewide}} & \multicolumn{2}{c}{\hdtop{Setting}} & \multicolumn{2}{c}{\hdtop{Road system}} & \multicolumn{2}{c}{\hdtop{Year}} \\
\headrow \hdsub{Factor} & \hdsub{Crashes} & \hdsub{Interval} & \hdsub{Rate} & \hdsub{Rural} & \hdsub{Urban} & \hdsub{State} & \hdsub{Local} & \hdsub{2019} & \hdsub{2021} \\
\midrule
Alcohol & 37,061 & 26,713 to 51,177 & \cellbar[discver]{1.00}{7.4} & 8.4 & 7.0 & 6.9 & 8.0 & 7.7 & 8.5 \\
Hydroplane & 34,198 & 30,463 to 38,845 & \cellbar[discver]{0.92}{6.8} & \best{8.6} & 6.2 & \best{8.6} & 4.7 & 7.0 & 7.7 \\
Wrong-way & 33,418 & 16,391 to 51,926 & \cellbar[discver]{0.90}{6.7} & 6.3 & 6.8 & 5.9 & 7.5 & 6.8 & 6.7 \\
Phone use$^{\dagger}$ & 26,754 & 22,611 to 34,425 & \cellbar[discver]{0.72}{5.3} & 5.7 & 5.2 & 5.3 & 5.4 & 4.8 & 5.5 \\
Medical episode & 25,979 & 16,476 to 31,405 & \cellbar[discver]{0.70}{5.2} & 5.5 & 5.1 & 5.1 & 5.3 & 4.4 & 5.2 \\
Fatigue & 22,690 & 19,095 to 24,206 & \cellbar[discver]{0.61}{4.5} & \best{7.4} & 3.5 & \best{5.3} & 3.5 & 5.2 & 4.8 \\
Animal & 15,120 & 11,099 to 16,935 & \cellbar[discver]{0.41}{3.0} & \best{5.9} & 2.0 & 2.6 & \best{3.5} & 3.1 & 3.0 \\
Drugs$^{\dagger}$ & 10,719 & 6,780 to 12,037 & \cellbar[discver]{0.29}{2.1} & \best{3.1} & 1.8 & 1.9 & 2.4 & 2.6 & 2.1 \\
Unbelted$^{\dagger}$ & 8,556 & 4,777 to 13,209 & \cellbar[discver]{0.23}{1.7} & 1.3 & \best{1.8} & 1.5 & 2.0 & 2.0 & 1.7 \\
\bottomrule
\end{tabular}
\tablenote{Crashes are multiwave estimates over all crashes in $U$ with 95 percent intervals, and rates are per 1,000 crashes in the domain, drawn as bars against the highest statewide rate. Bold marks the higher rate of a pair when it exceeds the lower by more than a quarter of the higher, computed on the printed values. County results for every factor are in the released county table, and Figure~\ref{fig:atlas}(b) shows them for phone use. In both, a county is pooled when its crash count times the statewide rate falls below 10. $^{\dagger}$Pooled recalibration map.}
\end{table}

\begin{figure}[t]
\centering
\includegraphics[width=\textwidth]{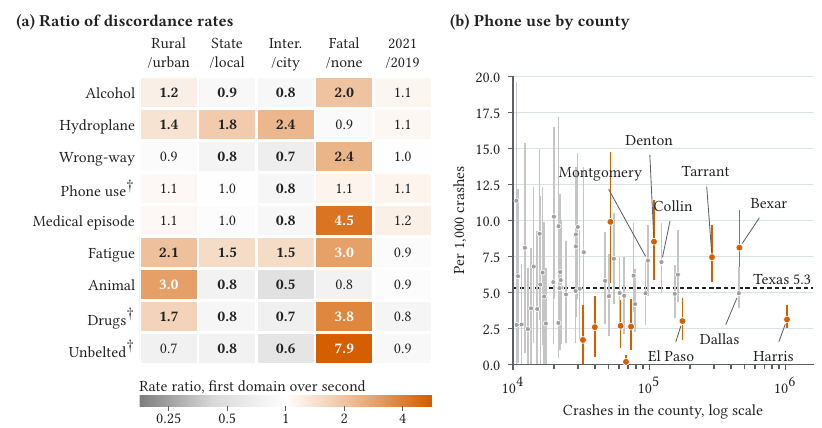}
\caption{Where confirmed discordance concentrates. Panel (a) gives, for each pair of domains, the ratio of their discordance rates per 1,000 crashes, in bold where the paired 95 percent interval of the ratio excludes one. The pairs are rural against urban, the state system against local roads, interstates against city streets, fatal crashes against crashes without injury, and 2021 against 2019. Panel (b) gives, for phone use, the rate of each county with at least \CountyMinCrashes{} crashes that passes the suppression rule, with its 95 percent interval, in vermilion where the interval excludes the statewide rate. $^{\dagger}$Pooled recalibration map.}\label{fig:atlas}
\end{figure}

Figure~\ref{fig:atlas} gives the contrasts between pairs of domains for each factor. Discordance for animals is \ConRuralAnimal{} times as high in rural as in urban crashes (95 percent interval \ConRuralAnimalLo{} to \ConRuralAnimalHi{}), fatigue, drugs, hydroplaning, and alcohol are also higher in rural crashes, and hydroplaning and fatigue are higher on the state highway system than on local roads. Fatal crashes carry higher discordance rates than crashes without injury for alcohol, drugs, medical episodes, fatigue, wrong-way driving, and unbelted occupants, but not for hydroplaning, animals, or phone use. The paired ratio between 2021 and 2019 includes one for every factor, so no discordance rate changed significantly across the 2020 change of the crash report. Counties differ more than the statewide splits, with phone-use discordance significantly below the statewide rate in Harris and El Paso, at about three per 1,000 crashes, and significantly above it in Bexar, Tarrant, Denton, and Smith, at between about seven and ten.

\begin{figure}[t]
\centering
\includegraphics[width=\textwidth]{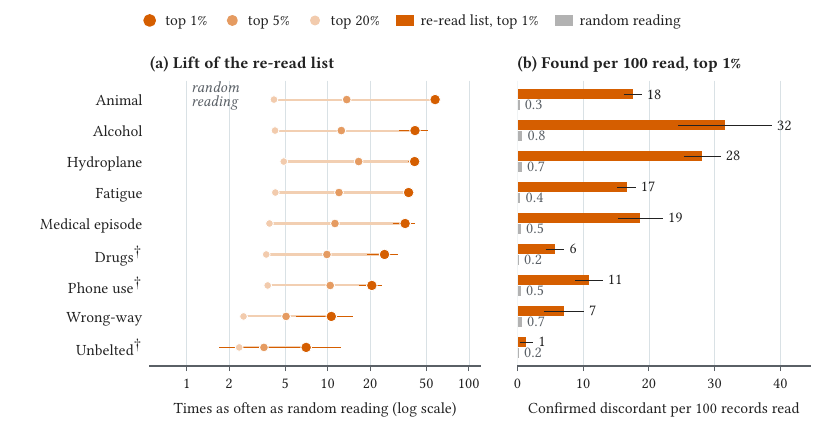}
\caption{The re-read list, measured on the second-wave sample. Panel (a) gives, for each factor, the lift over random reading at the top one, five, and twenty percent of the unread narratives, with the 95 percent interval at one percent. Panel (b) gives the records confirmed discordant per hundred read from the top one percent of the list, with its interval, and in a random reading of the same length. $^{\dagger}$Pooled recalibration map.}\label{fig:reread}
\end{figure}

The re-read list finds confirmed discordance several times as often as random reading, which answers the list part of RQ4. For the top one percent of unread narratives the lift over random reading is \LiftOneAnimal{} for animals, \LiftOneHydroplane{} for hydroplaning, \LiftOneAlcohol{} for alcohol, and \LiftOneMedical{} for medical episodes (Figure~\ref{fig:reread}), and for alcohol \PrecOneAlcohol{} percent of the listed records are confirmed discordant against \RandPrecAlcohol{} percent at random. The lift is smaller where the table reader is weakest, \LiftOneWrongWay{} for wrong-way driving and \LiftOneUnbelted{} for unbelted occupants, and it stays similar when the list is restricted to injury-or-fatal crashes.

\begin{figure}[t]
\centering
\includegraphics[width=\textwidth]{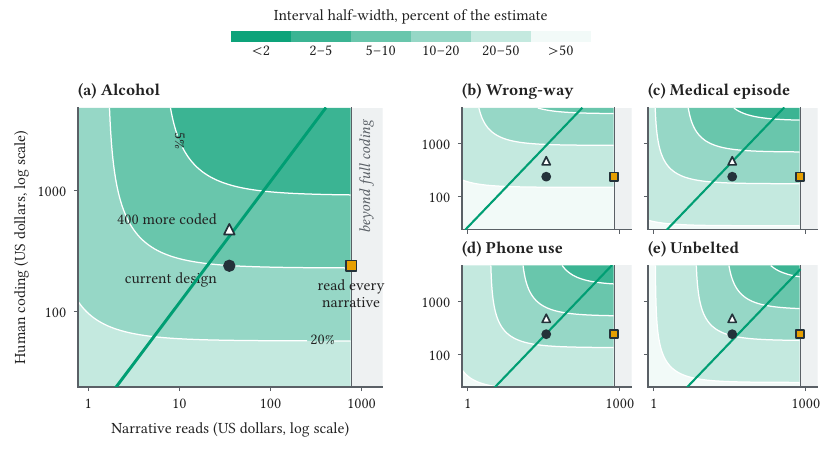}
\caption{Iso-precision maps of the reading budget. Each panel spans spending on narrative reads against spending on human-coded narratives, shaded by the half-width of the 95 percent interval relative to the estimate. The green curve is the cheapest path to each precision, and the markers are the current design (dot), \LabelRound{} more human-coded narratives (triangle), and reading every narrative (square), and the grey band lies beyond full coding. The cost of a human-coded narrative is a planning value of \HumanCostUsd{} dollars, six judgments at ten cents each from the coder throughput of the phase-2 human tier.}\label{fig:budget}
\end{figure}

A future budget should buy human judgments rather than more narrative reads, which answers the budget part of RQ4. Reading every narrative in $U$, about \CostFullCoding{} dollars at the vendor's price, would narrow the alcohol interval by \FullGainAlcohol{} percent, while one further round of \LabelRound{} human-coded narratives would narrow it by \PlusGainAlcohol{} percent (Figure~\ref{fig:budget}). Every factor gains more from that round. Over the fifteen factors the root mean square relative half-width falls from \JointRmsNow{} percent now to \JointRmsFull{} percent after reading every remaining narrative and to \JointRmsPlus{} percent after the round. The round buys more precision per dollar than full reading at any human cost below \JointBreakeven{} dollars per narrative over the fifteen factors, and below \BreakevenMin{} dollars for every factor, against a planning cost of \HumanCostUsd{} dollars. Refitting the maps on resampled human tiers of two, three, and four hundred narratives gave a median exponent of \ScaleAlphaMedian{} over the fifteen factors for the decline of the recalibration variance. That is close to the exponent of one that the joint projection assumes, although single factors range from \ScaleAlphaMin{} to \ScaleAlphaMax{}. A human tier stratified on the disagreement between the two readers would not have improved on the stratification by the narrative reader's probability that drew the phase-2 tier, with a variance ratio of \RetroRatio{} (95 percent interval \RetroRatioLo{} to \RetroRatioHi{}, Appendix~\ref{app:recal}).

Kumo Tabular small is an adequate reader of the crash table and far faster than the most accurate one, which answers RQ5. On the same \BenchTest{} test rows of one wave-1 fold, TabPFN 3.5 discriminated best at \BenchCtx{} context rows, with a mean flag AUC of \AucFlagPFNNew{} against \AucFlagKumo{} for Kumo Tabular small, and the paired bootstrap favors it on \NBetterPFNNew{} of the \NBenchTargets{} targets and Kumo Tabular small on \NWorsePFNNew{} (Table~\ref{tab:benchmark}). Kumo Tabular small in turn leads TabPFN 3, TabICLv2, and CatBoost trained on the same context, which are never significantly better and are worse on \NWorsePFNOld{}, \NWorseTabICL{}, and \NWorseCatBoost{} targets in comparisons not adjusted for multiplicity. Over all five folds it discriminated at least as well as a gradient-boosted tree tuned per target on every one of the \CfNTargets{} targets, with a mean flag AUC of \CfAucKumo{} against \CfAucBooster{}. Against human labels its mean AUC was \HumAucKumoOne{} against \HumAucBoosterOne{} over the fifteen factors of the phase-2 tier and \HumAucKumoTwo{} against \HumAucBoosterTwo{} over the fourteen factors of the second with enough positives. CatBoost trained on the whole pool, twice the largest context, is significantly better than Kumo Tabular small only for witness citations.

The price of the advantage of TabPFN 3.5 is speed. With the default inference settings of each library, Kumo Tabular small scored \SpeedRatioPFN{} times as many rows per second, so one pass over all \NAll{} crashes takes about \PassMinKumo{} minutes against \PassHoursPFN{} hours, and the cross-fitted design makes \Kfolds{} passes per target. The gap in discrimination matters little for the estimator, whose gain depends on the squared correlation of the table prediction with the recalibrated narrative probability, \RhoSqKumo{} for Kumo Tabular small and \RhoSqPFNNew{} for TabPFN 3.5. Kumo Tabular small was fixed as the production reader before this comparison, which therefore tests its adequacy rather than selecting it.

\begin{table}[t]
\caption{The table readers on the same test rows.}\label{tab:benchmark}
\tablestyle
\begin{tabular}{P{0.22}R{0.085}R{0.07}R{0.105}R{0.07}R{0.065}R{0.07}R{0.175}R{0.06}}
\toprule
\headrow \hd{} & \hd{} & \multicolumn{2}{c}{\hdtop{AUC}} & \multicolumn{2}{c}{\hdtop{Calibration}} & \hd{} & \multicolumn{2}{c}{\hdtop{Cost}} \\
\headrow \hdsub{Reader} & \hdsub{Context} & \hdsub{Flags} & \hdsub{Discordance} & \hdsub{Error} & \hdsub{Slope} & \hdsub{$\rho^2$} & \hdsub{Rows per s} & \hdsub{GPU GB} \\
\midrule
Kumo Tabular, small & 60,000 & 0.865 & 0.759 & 0.002 & 1.01 & 0.279 & \cellbar[kumoblue]{0.76}{8,393} & 12.1 \\
Kumo Tabular, medium & 30,000 & 0.861 & 0.741 & 0.002 & 1.01 & 0.280 & \cellbar[kumoblue]{0.58}{2,872} & 66.6 \\
Kumo Tabular, large & 10,000 & 0.853 & 0.718 & 0.003 & 1.03 & 0.269 & \cellbar[kumoblue]{0.62}{3,632} & 45.4 \\
TabPFN 3.5 & 60,000 & \best{0.872} & \best{0.772} & 0.002 & 1.02 & \best{0.286} & \cellbar[kumoblue]{0.29}{556} & 3.0 \\
TabPFN 3 & 60,000 & 0.859 & 0.746 & 0.004 & 0.95 & 0.278 & \cellbar[kumoblue]{0.40}{1,004} & 1.7 \\
TabICLv2 & 60,000 & 0.850 & 0.709 & 0.005 & 0.90 & 0.275 & \cellbar[kumoblue]{0.79}{9,899} & 9.1 \\
CatBoost, same context & 60,000 & 0.854 & 0.724 & 0.002 & 1.01 & 0.272 & \cellbar[kumoblue]{1.00}{34,198} & \cellnote{CPU} \\
CatBoost, whole pool & 120,000 & 0.858 & 0.737 & 0.002 & 0.98 & 0.276 & \cellbar[kumoblue]{1.00}{33,600} & \cellnote{CPU} \\
\bottomrule
\end{tabular}
\tablenote{All readers predict the same \BenchTest{} test rows of one wave-1 fold from nested contexts drawn from the other folds. AUC, calibration error, and slope are means over \BenchFlags{} flag targets, and Discordance is the mean AUC over \BenchDisc{} discordance targets scored on crashes whose coded field is false. $\rho^2$ is the mean squared correlation of the prediction with the recalibrated narrative probability over all \NBenchTargets{} targets. Calibration is measured against the narrative reader's flag. Medium and large Kumo Tabular are shown at the largest context of their grid, which GPU memory limited (medium used \GpuKumoMediumThirty{} GB at 30,000 rows). Bold marks the highest AUC and $\rho^2$ at printed precision. Rows per second are test rows scored per second, drawn as a bar on a log scale from 100 to the fastest reader, measured on one A100 GPU, except CatBoost, which was trained and run on the CPU. Single-fold AUCs differ from the cross-fitted AUCs of Table~\ref{tab:predictability} for rare targets, and the booster of that table differs from CatBoost here in library and settings, so the two need not agree.}
\end{table}

\section{Discussion}\label{sec:discussion}

The central result is that the two views of a crash record are complementary in a way that can be measured factor by factor and turned into estimates whose uncertainty is stated tier by tier. Where a coded field exists and its evidence is routine, the coded record predicts the narrative reading closely, and where the evidence depends on what an officer observes or hears, the narrative carries most of the information. The Arkansas benchmark of six frontier language models against official coding found that differences across attributes exceeded differences across models~\cite{bharati2026benchmarking}. The present results give that observation a quantitative form at the scale of a state, since predictability and narrative surplus vary far more across factors than the change of table reader moves them.

The factor-level counts agree with what the under-reporting literature leads one to expect and add magnitudes with intervals for one state. Phone use is the clearest case, with about twice as many injury crashes documented in the narrative as coded in the field. The national distraction note lists reliance on self-report, on the post-crash investigation, and on report forms among the reasons that distraction may be under-reported~\cite{nhtsa2025distracted}. Fatigue points the same way, though far less strongly than the finding that a proxy definition identifies nearly three times as many sleep-related crash involvements as police reports~\cite{filtness2017sleep}. Alcohol and drugs are the coded factors whose field records about as many injury crashes as the narrative or more. For alcohol this is consistent with the Texas instruction that specimen results always go in the coded field~\cite{txdot2025instructions}, under which the field can record alcohol that the narrative need not repeat.

Restraint use shows why discordance must be confirmed before it is counted. The coded field marks an unrestrained driver in about six times as many injury crashes as the narrative states that someone was unbelted. Human coders confirmed about one in seven of the \ConfFalseUnbeltedN{} judged narrative flags that the field did not carry, against about nine in ten for phone use and fatigue. A comparison against crash investigators found that police belt reports had high sensitivity and low specificity~\cite{schiff2004comparison}, which describes errors in the field, whereas the present comparison sets a driver-level field against an occupant-level narrative statement. A blind review of speeding designations found that only about half of the designated crashes described speeding in the narrative~\cite{fitzpatrick2017investigation}, and a Massachusetts study found discrepancies between narratives and coded impaired-driving fields in both directions~\cite{das2026revealing}. A discordance count without human confirmation would therefore overstate the narrative surplus for some factors several times over.

The design results speak to the literature on inference with model predictions. Prediction-powered inference and its power-tuned version show that a good predictor narrows intervals without biasing them~\cite{angelopoulos2023prediction,angelopoulos2026ppi}, and the present table reader does that for the design tier. The recalibration tier, however, carries most of the uncertainty for every factor, a limit that no table reader could lift, so the table reader contributes most by targeting the second wave and ranking the re-read list. Design-based supervised learning with noisy human labels showed that the quality of the human tier shapes what the estimator can deliver~\cite{chew2026design}. The present variance split makes the same point in budget terms. The second human tier shows its other side, since a few hundred specialist-coded narratives drawn with recorded probabilities agreed with every estimate within its sampling margin, so a model-assisted count can be audited by design at modest cost.

The table-reader evaluation adds a domain check to the benchmark record of the new foundation models. Kumo Tabular was released with a vendor-reported top rank on public tabular benchmarks~\cite{qu2026nvidia}, yet on this crash table TabPFN 3.5~\cite{jager2026tabpfn} discriminated slightly better, so a benchmark rank need not carry over to a particular domain table. Kumo Tabular small matched or exceeded a tuned booster on every target over all five folds, which supports on a real domain table the finding that in-context readers match or exceed tuned boosters without training~\cite{hollmann2025accurate,qu2026tabiclv2}. It also extends to in-context readers the earlier result that a deep tabular model can outperform an ensemble in overall accuracy on crash data~\cite{rafe2024comparative}. Because the estimator gains from the squared correlation of the reader with the recalibrated narrative probability, which is nearly equal across readers, the choice among them turns on throughput, hardware, and license rather than on discrimination.

The outputs translate into operational steps for a safety office. The atlas shows where a review of whether discordance reflects reporting practice, procedure, or omission would start. The re-read list ranks unread records by the probability that their narrative states a factor that the field omits, and the allocation rule turns the measured variance components into a purchase decision for the next reading budget. The method should transfer to any state whose crash reports carry a narrative and coded fields aligned with the voluntary Model Minimum Uniform Crash Criteria~\cite{nhtsa2017mmucc}. It needs a coded record, narratives, a probabilistic reader for each view, and a human tier that a probability sample of a few hundred specialist-coded narratives can check by design.

\section{Conclusion}\label{sec:conclusion}

This paper composed two decision models and a human tier into one estimation system for the two views of a crash record and applied it to the \NAll{} Texas crashes of 2017 to 2025, with estimates over the \NU{} crashes that carry a usable narrative. The coded record predicted the narrative where its evidence is routine and missed it where only the narrative carries the information, which answers RQ1. For hydroplaning, medical episodes, fatigue, animals, and phone use the narrative documents more injury crashes than the coded field records, while the field records about as many or more for alcohol and drugs and far more for unbelted occupants. A second human tier drawn with recorded probabilities agreed with every count within its sampling margin, which answers RQ2. The atlas located confirmed discordance by setting, road system, and county, which answers RQ3. The re-read list found confirmed discordance \LiftOneUnbelted{} to \LiftOneAnimal{} times as often as random reading at the top one percent. At the planning cost of human coding, the next reading budget buys more precision as human judgments than as narrative reads for every factor, which answers RQ4. Kumo Tabular small matched or exceeded a tuned booster and came close to TabPFN 3.5 while scoring rows \SpeedRatioPFN{} times faster, which answers RQ5. Two calibrated readers of different views, joined by a sampling design, thus give a safety office population counts, a map, a list, and a budget, each with its uncertainty stated tier by tier.

Three conditions set the scope of the study. The narrative reader is a closed commercial model whose training data are not disclosed, so its readings were pinned to one version, audited against two human tiers, and recalibrated rather than inspected. The extrapolations to crashes without a usable narrative are model-based by construction and are reported apart from every estimate. The study covers one state, whose crash report follows common minimum criteria that other states share.

These conditions point directly to the next steps for this line of work. Repeating the design in other states would show how much of the discordance is Texas practice, and repeating it across versions of both readers would show how stable the estimates are when the readers change. Reading the narratives of the crashes without coordinates after a cleaning pass would bring the extrapolated group into the measured population, and a larger human tier drawn by the allocation rule would narrow the intervals that the recalibration tier dominates. A relational table reader that sees the crash, unit, and person tables jointly is the natural extension of the coded view.

\section*{Data and code availability}

The crash records are the Texas Crash Records Information System extract for 2017 to 2025, held under a data agreement with the Texas Department of Transportation that does not permit redistribution. Aggregate CRIS data are public at \url{https://cris.dot.state.tx.us/public/Query/app/home}, and bulk extracts are requested at \url{https://www.txdot.gov/apps-cg/crash_records/form.htm}. The analysis and plotting code, the question schemas, the coder guide, the design summaries, and the aggregate result tables are at \url{https://github.com/pozapas/kumo-jev-crash-records}.

\section*{Ethics}

The university's institutional review board determined that this secondary analysis of administrative records is not human subjects research. The data owner replaced names, telephone numbers, and other identifiers with placeholders before transfer, only these cleaned narratives were sent to the narrative reader, and no narrative is reproduced in the paper.

\section*{Funding}

This work received no specific funding from any agency, company, or foundation. TabPFN was used under its evaluation license.

\section*{Author contributions}

Amir Rafe conceived the study, designed the methods, carried out the analysis, and wrote the manuscript. Subasish Das contributed to the study design, provided the data, and reviewed and edited the manuscript.

\section*{Competing interests}

The authors declare that they have no competing interests related to this work. Neither TypeSafe, the developer of Jev, nor NVIDIA, the developer of Kumo Tabular, provided credits, early access, or review of this work, and the authors have no financial or advisory relationship with either company.

\FloatBarrier
\bibliographystyle{ACM-Reference-Format}
\bibliography{refs}


\begin{thebibliography}{53}


\ifx \showCODEN    \undefined \def \showCODEN     #1{\unskip}     \fi
\ifx \showISBNx    \undefined \def \showISBNx     #1{\unskip}     \fi
\ifx \showISBNxiii \undefined \def \showISBNxiii  #1{\unskip}     \fi
\ifx \showISSN     \undefined \def \showISSN      #1{\unskip}     \fi
\ifx \showLCCN     \undefined \def \showLCCN      #1{\unskip}     \fi
\ifx \shownote     \undefined \def \shownote      #1{#1}          \fi
\ifx \showarticletitle \undefined \def \showarticletitle #1{#1}   \fi
\ifx \showURL      \undefined \def \showURL       {\relax}        \fi
\providecommand\bibfield[2]{#2}
\providecommand\bibinfo[2]{#2}
\providecommand\natexlab[1]{#1}
\providecommand\showeprint[2][]{arXiv:#2}

\bibitem[Angelopoulos et~al\mbox{.}(2023)]%
        {angelopoulos2023prediction}
\bibfield{author}{\bibinfo{person}{Anastasios~N. Angelopoulos},
  \bibinfo{person}{Stephen Bates}, \bibinfo{person}{Clara Fannjiang},
  \bibinfo{person}{Michael~I. Jordan}, {and} \bibinfo{person}{Tijana Zrnic}.}
  \bibinfo{year}{2023}\natexlab{}.
\newblock \showarticletitle{Prediction-powered inference}.
\newblock \bibinfo{journal}{\emph{Science}} \bibinfo{volume}{382},
  \bibinfo{number}{6671} (\bibinfo{year}{2023}), \bibinfo{pages}{669--674}.
\newblock
\href{https://doi.org/10.1126/science.adi6000}{doi:\nolinkurl{10.1126/science.adi6000}}


\bibitem[Angelopoulos et~al\mbox{.}(2026)]%
        {angelopoulos2026ppi}
\bibfield{author}{\bibinfo{person}{Anastasios~N. Angelopoulos},
  \bibinfo{person}{John~C. Duchi}, {and} \bibinfo{person}{Tijana Zrnic}.}
  \bibinfo{year}{2026}\natexlab{}.
\newblock \showarticletitle{{PPI++}: Efficient prediction-powered inference}.
\newblock \bibinfo{journal}{\emph{The Annals of Applied Statistics}}
  \bibinfo{volume}{20}, \bibinfo{number}{3} (\bibinfo{year}{2026}),
  \bibinfo{pages}{2235--2252}.
\newblock
\showeprint[arxiv]{2311.01453}
\href{https://doi.org/10.1214/26-aoas2215}{doi:\nolinkurl{10.1214/26-aoas2215}}


\bibitem[Arteaga and Park(2025)]%
        {arteaga2025large}
\bibfield{author}{\bibinfo{person}{Cristian Arteaga} {and}
  \bibinfo{person}{JeeWoong Park}.} \bibinfo{year}{2025}\natexlab{}.
\newblock \showarticletitle{A large language model framework to uncover
  underreporting in traffic crashes}.
\newblock \bibinfo{journal}{\emph{Journal of Safety Research}}
  \bibinfo{volume}{92} (\bibinfo{year}{2025}), \bibinfo{pages}{1--13}.
\newblock
\href{https://doi.org/10.1016/j.jsr.2024.11.009}{doi:\nolinkurl{10.1016/j.jsr.2024.11.009}}


\bibitem[Bhagat et~al\mbox{.}(2025)]%
        {bhagat2025accuracy}
\bibfield{author}{\bibinfo{person}{Sudesh Bhagat}, \bibinfo{person}{Ibne~Farabi
  Shihab}, \bibinfo{person}{Jonathan Wood}, {and} \bibinfo{person}{Anuj
  Sharma}.} \bibinfo{year}{2025}\natexlab{}.
\newblock \showarticletitle{Accuracy is Not Agreement: {Expert-Aligned}
  Evaluation of Crash Narrative Classification Models}. In
  \bibinfo{booktitle}{\emph{2025 IEEE 28th International Conference on
  Intelligent Transportation Systems (ITSC)}}. \bibinfo{publisher}{IEEE},
  \bibinfo{pages}{3236--3243}.
\newblock
\showeprint[arxiv]{2504.13068}
\href{https://doi.org/10.1109/itsc60802.2025.11423032}{doi:\nolinkurl{10.1109/itsc60802.2025.11423032}}


\bibitem[Bharati et~al\mbox{.}(2026)]%
        {bharati2026benchmarking}
\bibfield{author}{\bibinfo{person}{Sudhir Bharati}, \bibinfo{person}{Rajendra
  K~C Khatri}, {and} \bibinfo{person}{Sudip Bharati}.}
  \bibinfo{year}{2026}\natexlab{}.
\newblock \bibinfo{title}{Benchmarking Frontier Large Language Models Against
  Official Crash Database Coding Using Police Crash Narratives}.
\newblock \bibinfo{howpublished}{arXiv preprint arXiv:2607.29064}.
\newblock
\showeprint[arxiv]{2607.29064}~[cs.LG]
\urldef\tempurl%
\url{https://arxiv.org/abs/2607.29064}
\showURL{%
\tempurl}


\bibitem[Breslow et~al\mbox{.}(2009)]%
        {breslow2009improved}
\bibfield{author}{\bibinfo{person}{Norman~E. Breslow}, \bibinfo{person}{Thomas
  Lumley}, \bibinfo{person}{Christie~M. Ballantyne}, \bibinfo{person}{Lloyd~E.
  Chambless}, {and} \bibinfo{person}{Michal Kulich}.}
  \bibinfo{year}{2009}\natexlab{}.
\newblock \showarticletitle{Improved {Horvitz–Thompson} Estimation of Model
  Parameters from Two-phase Stratified Samples: Applications in Epidemiology}.
\newblock \bibinfo{journal}{\emph{Statistics in Biosciences}}
  \bibinfo{volume}{1}, \bibinfo{number}{1} (\bibinfo{year}{2009}),
  \bibinfo{pages}{32--49}.
\newblock
\href{https://doi.org/10.1007/s12561-009-9001-6}{doi:\nolinkurl{10.1007/s12561-009-9001-6}}


\bibitem[Brier(1950)]%
        {brier1950verification}
\bibfield{author}{\bibinfo{person}{Glenn~W. Brier}.}
  \bibinfo{year}{1950}\natexlab{}.
\newblock \showarticletitle{Verification of Forecasts Expressed in Terms of
  Probability}.
\newblock \bibinfo{journal}{\emph{Monthly Weather Review}}
  \bibinfo{volume}{78}, \bibinfo{number}{1} (\bibinfo{year}{1950}),
  \bibinfo{pages}{1--3}.
\newblock
\href{https://doi.org/10.1175/1520-0493(1950)078<0001:vofeit>2.0.co;2}{doi:\nolinkurl{10.1175/1520-0493(1950)078<0001:vofeit>2.0.co;2}}


\bibitem[Chew and Williams(2026)]%
        {chew2026design}
\bibfield{author}{\bibinfo{person}{Robert Chew} {and}
  \bibinfo{person}{Matthew~R. Williams}.} \bibinfo{year}{2026}\natexlab{}.
\newblock \bibinfo{title}{{Design-Based} Supervised Learning with Noisy Human
  Labels}.
\newblock \bibinfo{howpublished}{arXiv preprint arXiv:2607.15455}.
\newblock
\showeprint[arxiv]{2607.15455}~[stat.ML]
\urldef\tempurl%
\url{https://arxiv.org/abs/2607.15455}
\showURL{%
\tempurl}


\bibitem[Das et~al\mbox{.}(2026)]%
        {das2026revealing}
\bibfield{author}{\bibinfo{person}{Niloy Das}, \bibinfo{person}{Jee~Woong
  Park}, {and} \bibinfo{person}{Tilak Barua}.} \bibinfo{year}{2026}\natexlab{}.
\newblock \bibinfo{title}{Revealing Contextual Discrepancies between Structured
  and Narrative Crash Report Data in {DUI-Related} Crashes}.
\newblock \bibinfo{howpublished}{SSRN preprint}.
\newblock
\href{https://doi.org/10.2139/ssrn.6519521}{doi:\nolinkurl{10.2139/ssrn.6519521}}


\bibitem[Egami et~al\mbox{.}(2023)]%
        {egami2023using}
\bibfield{author}{\bibinfo{person}{Naoki Egami}, \bibinfo{person}{Musashi
  Hinck}, \bibinfo{person}{Brandon Stewart}, {and} \bibinfo{person}{Hanying
  Wei}.} \bibinfo{year}{2023}\natexlab{}.
\newblock \showarticletitle{Using Imperfect Surrogates for Downstream
  Inference: Design-based Supervised Learning for Social Science Applications
  of Large Language Models}. In \bibinfo{booktitle}{\emph{Advances in Neural
  Information Processing Systems 36}}. \bibinfo{publisher}{Neural Information
  Processing Systems Foundation, Inc. (NeurIPS)},
  \bibinfo{pages}{68589--68601}.
\newblock
\showeprint[arxiv]{2306.04746}
\href{https://doi.org/10.52202/075280-3000}{doi:\nolinkurl{10.52202/075280-3000}}


\bibitem[Elvik and Mysen(1999)]%
        {elvik1999incomplete}
\bibfield{author}{\bibinfo{person}{Rune Elvik} {and} \bibinfo{person}{Anne
  Mysen}.} \bibinfo{year}{1999}\natexlab{}.
\newblock \showarticletitle{Incomplete Accident Reporting: {Meta-Analysis} of
  Studies Made in 13 Countries}.
\newblock \bibinfo{journal}{\emph{Transportation Research Record: Journal of
  the Transportation Research Board}} \bibinfo{volume}{1665},
  \bibinfo{number}{1} (\bibinfo{year}{1999}), \bibinfo{pages}{133--140}.
\newblock
\href{https://doi.org/10.3141/1665-18}{doi:\nolinkurl{10.3141/1665-18}}


\bibitem[Erickson et~al\mbox{.}(2025)]%
        {erickson2025tabarena}
\bibfield{author}{\bibinfo{person}{Nick Erickson}, \bibinfo{person}{Lennart
  Purucker}, \bibinfo{person}{Andrej Tschalzev}, \bibinfo{person}{David
  Holzmüller}, \bibinfo{person}{Prateek Desai}, \bibinfo{person}{David
  Salinas}, {and} \bibinfo{person}{Frank Hutter}.}
  \bibinfo{year}{2025}\natexlab{}.
\newblock \showarticletitle{{TabArena}: A Living Benchmark for Machine Learning
  on Tabular Data}. In \bibinfo{booktitle}{\emph{Advances in Neural Information
  Processing Systems 38}}. \bibinfo{publisher}{Neural Information Processing
  Systems Foundation, Inc. (NeurIPS)}, \bibinfo{pages}{17285--17350}.
\newblock
\showeprint[arxiv]{2506.16791}
\href{https://doi.org/10.52202/085713-0519}{doi:\nolinkurl{10.52202/085713-0519}}


\bibitem[Filtness et~al\mbox{.}(2017)]%
        {filtness2017sleep}
\bibfield{author}{\bibinfo{person}{A.J. Filtness}, \bibinfo{person}{K.A.
  Armstrong}, \bibinfo{person}{A. Watson}, {and} \bibinfo{person}{S.S. Smith}.}
  \bibinfo{year}{2017}\natexlab{}.
\newblock \showarticletitle{Sleep-related crash characteristics: Implications
  for applying a fatigue definition to crash reports}.
\newblock \bibinfo{journal}{\emph{Accident Analysis \& Prevention}}
  \bibinfo{volume}{99} (\bibinfo{year}{2017}), \bibinfo{pages}{440--444}.
\newblock
\href{https://doi.org/10.1016/j.aap.2015.11.024}{doi:\nolinkurl{10.1016/j.aap.2015.11.024}}


\bibitem[Fitzpatrick et~al\mbox{.}(2017)]%
        {fitzpatrick2017investigation}
\bibfield{author}{\bibinfo{person}{Cole~D. Fitzpatrick},
  \bibinfo{person}{Saritha Rakasi}, {and} \bibinfo{person}{Michael~A.
  Knodler}.} \bibinfo{year}{2017}\natexlab{}.
\newblock \showarticletitle{An investigation of the speeding-related crash
  designation through crash narrative reviews sampled via logistic regression}.
\newblock \bibinfo{journal}{\emph{Accident Analysis \& Prevention}}
  \bibinfo{volume}{98} (\bibinfo{year}{2017}), \bibinfo{pages}{57--63}.
\newblock
\href{https://doi.org/10.1016/j.aap.2016.09.017}{doi:\nolinkurl{10.1016/j.aap.2016.09.017}}


\bibitem[Grinsztajn et~al\mbox{.}(2026)]%
        {grinsztajn2026tabpfn}
\bibfield{author}{\bibinfo{person}{Léo Grinsztajn}, \bibinfo{person}{Klemens
  Flöge}, \bibinfo{person}{Oscar Key}, \bibinfo{person}{Felix Birkel},
  \bibinfo{person}{Philipp Jund}, \bibinfo{person}{Brendan Roof},
  \bibinfo{person}{Mihir Manium}, \bibinfo{person}{Shi~Bin Hoo},
  \bibinfo{person}{Magnus Bühler}, \bibinfo{person}{Anurag Garg},
  \bibinfo{person}{Dominik Safaric}, \bibinfo{person}{Jake Robertson},
  \bibinfo{person}{Benjamin Jäger}, \bibinfo{person}{Simone Alessi},
  \bibinfo{person}{Adrian Hayler}, \bibinfo{person}{Vladyslav Moroshan},
  \bibinfo{person}{Lennart Purucker}, \bibinfo{person}{Philipp Singer},
  \bibinfo{person}{Alan Arazi}, \bibinfo{person}{Julien Siems},
  \bibinfo{person}{Jan~Hendrik Metzen}, \bibinfo{person}{Georg Grab},
  \bibinfo{person}{Nick Erickson}, \bibinfo{person}{Siyuan Guo},
  \bibinfo{person}{Eliott Kalfon}, \bibinfo{person}{Simon Bing},
  \bibinfo{person}{David Salinas}, \bibinfo{person}{Clara Cornu},
  \bibinfo{person}{Lilly~Charlotte Wehrhahn}, \bibinfo{person}{Diana
  Kriuchkova}, \bibinfo{person}{Kursat Kaya}, \bibinfo{person}{Lydia Sidhoum},
  \bibinfo{person}{Marie Salmon}, \bibinfo{person}{Jerry Chen},
  \bibinfo{person}{Madelon Hulsebos}, \bibinfo{person}{Yann LeCun},
  \bibinfo{person}{Samuel Müller}, \bibinfo{person}{Bernhard Schölkopf},
  \bibinfo{person}{Sauraj Gambhir}, \bibinfo{person}{Noah Hollmann}, {and}
  \bibinfo{person}{Frank Hutter}.} \bibinfo{year}{2026}\natexlab{}.
\newblock \bibinfo{title}{{TabPFN-3}: Technical Report}.
\newblock \bibinfo{howpublished}{arXiv preprint arXiv:2605.13986}.
\newblock
\showeprint[arxiv]{2605.13986}~[cs.LG]
\urldef\tempurl%
\url{https://arxiv.org/abs/2605.13986}
\showURL{%
\tempurl}


\bibitem[Hollmann et~al\mbox{.}(2025)]%
        {hollmann2025accurate}
\bibfield{author}{\bibinfo{person}{Noah Hollmann}, \bibinfo{person}{Samuel
  Müller}, \bibinfo{person}{Lennart Purucker}, \bibinfo{person}{Arjun
  Krishnakumar}, \bibinfo{person}{Max Körfer}, \bibinfo{person}{Shi~Bin Hoo},
  \bibinfo{person}{Robin~Tibor Schirrmeister}, {and} \bibinfo{person}{Frank
  Hutter}.} \bibinfo{year}{2025}\natexlab{}.
\newblock \showarticletitle{Accurate predictions on small data with a tabular
  foundation model}.
\newblock \bibinfo{journal}{\emph{Nature}} \bibinfo{volume}{637},
  \bibinfo{number}{8045} (\bibinfo{year}{2025}), \bibinfo{pages}{319--326}.
\newblock
\href{https://doi.org/10.1038/s41586-024-08328-6}{doi:\nolinkurl{10.1038/s41586-024-08328-6}}


\bibitem[Hosseini et~al\mbox{.}(2023)]%
        {hosseini2023application}
\bibfield{author}{\bibinfo{person}{Parisa Hosseini},
  \bibinfo{person}{Seyedalireza Khoshsirat}, \bibinfo{person}{Mohammad
  Jalayer}, \bibinfo{person}{Subasish Das}, {and} \bibinfo{person}{Huaguo
  Zhou}.} \bibinfo{year}{2023}\natexlab{}.
\newblock \showarticletitle{Application of text mining techniques to identify
  actual wrong-way driving ({WWD}) crashes in police reports}.
\newblock \bibinfo{journal}{\emph{International Journal of Transportation
  Science and Technology}} \bibinfo{volume}{12}, \bibinfo{number}{4}
  (\bibinfo{year}{2023}), \bibinfo{pages}{1038--1051}.
\newblock
\href{https://doi.org/10.1016/j.ijtst.2022.12.002}{doi:\nolinkurl{10.1016/j.ijtst.2022.12.002}}


\bibitem[Hu and Ghelichi(2026)]%
        {hu2026noise}
\bibfield{author}{\bibinfo{person}{James Hu} {and} \bibinfo{person}{Mahdi
  Ghelichi}.} \bibinfo{year}{2026}\natexlab{}.
\newblock \bibinfo{title}{Noise Immunity in {In-Context} Tabular Learning: An
  Empirical Robustness Analysis of {TabPFN's} Attention Mechanisms}.
\newblock \bibinfo{howpublished}{arXiv preprint arXiv:2604.04868}.
\newblock
\showeprint[arxiv]{2604.04868}~[cs.LG]
\urldef\tempurl%
\url{https://arxiv.org/abs/2604.04868}
\showURL{%
\tempurl}


\bibitem[Ibrahim and Zaki(2026)]%
        {ibrahim2026evaluating}
\bibfield{author}{\bibinfo{person}{Hazem Ibrahim} {and} \bibinfo{person}{Yasir
  Zaki}.} \bibinfo{year}{2026}\natexlab{}.
\newblock \bibinfo{title}{Evaluating Decision Models for Text Annotation in
  Computational Social Science}.
\newblock \bibinfo{howpublished}{arXiv preprint arXiv:2609.24574}.
\newblock
\showeprint[arxiv]{2609.24574}~[cs.CL]
\urldef\tempurl%
\url{https://arxiv.org/abs/2609.24574}
\showURL{%
\tempurl}


\bibitem[Imprialou and Quddus(2019)]%
        {imprialou2019crash}
\bibfield{author}{\bibinfo{person}{Marianna Imprialou} {and}
  \bibinfo{person}{Mohammed Quddus}.} \bibinfo{year}{2019}\natexlab{}.
\newblock \showarticletitle{Crash data quality for road safety research:
  Current state and future directions}.
\newblock \bibinfo{journal}{\emph{Accident Analysis \& Prevention}}
  \bibinfo{volume}{130} (\bibinfo{year}{2019}), \bibinfo{pages}{84--90}.
\newblock
\href{https://doi.org/10.1016/j.aap.2017.02.022}{doi:\nolinkurl{10.1016/j.aap.2017.02.022}}


\bibitem[Janstrup et~al\mbox{.}(2016)]%
        {janstrup2016understanding}
\bibfield{author}{\bibinfo{person}{Kira~H. Janstrup}, \bibinfo{person}{Sigal
  Kaplan}, \bibinfo{person}{Tove Hels}, \bibinfo{person}{Jens Lauritsen}, {and}
  \bibinfo{person}{Carlo~G. Prato}.} \bibinfo{year}{2016}\natexlab{}.
\newblock \showarticletitle{Understanding traffic crash under-reporting:
  Linking police and medical records to individual and crash characteristics}.
\newblock \bibinfo{journal}{\emph{Traffic Injury Prevention}}
  \bibinfo{volume}{17}, \bibinfo{number}{6} (\bibinfo{year}{2016}),
  \bibinfo{pages}{580--584}.
\newblock
\href{https://doi.org/10.1080/15389588.2015.1128533}{doi:\nolinkurl{10.1080/15389588.2015.1128533}}


\bibitem[Jäger et~al\mbox{.}(2026)]%
        {jager2026tabpfn}
\bibfield{author}{\bibinfo{person}{Benjamin Jäger}, \bibinfo{person}{Nick
  Erickson}, \bibinfo{person}{Léo Grinsztajn}, \bibinfo{person}{Felix Birkel},
  \bibinfo{person}{Klemens Flöge}, \bibinfo{person}{Oscar Key},
  \bibinfo{person}{Kürşat Kaya}, \bibinfo{person}{Jonas Kübler},
  \bibinfo{person}{Adèle Frankel}, \bibinfo{person}{Tobias Schröder},
  \bibinfo{person}{Anurag Garg}, \bibinfo{person}{Jan~Hendrik Metzen},
  \bibinfo{person}{David Salinas}, \bibinfo{person}{Simon Bing},
  \bibinfo{person}{Kristina Collins}, \bibinfo{person}{Tuana Çelik},
  \bibinfo{person}{Vahid Balazadeh}, \bibinfo{person}{Lydia Sidhoum},
  \bibinfo{person}{Tomás Pereda}, \bibinfo{person}{Brendan Roof},
  \bibinfo{person}{Andrej Tschalzev}, \bibinfo{person}{Siyuan Guo},
  \bibinfo{person}{Philipp Singer}, \bibinfo{person}{Lennart Purucker},
  \bibinfo{person}{Jake Robertson}, \bibinfo{person}{Marie Salmon},
  \bibinfo{person}{Philipp Jund}, \bibinfo{person}{Jerry Chen},
  \bibinfo{person}{Diana Kriuchkova}, \bibinfo{person}{Arthur Cahu},
  \bibinfo{person}{Eliott Kalfon}, \bibinfo{person}{Adrian Hayler},
  \bibinfo{person}{Georg Grab}, \bibinfo{person}{Vitor Monteiro},
  \bibinfo{person}{Lilly Wehrhahn}, \bibinfo{person}{Dominik Safaric},
  \bibinfo{person}{Clara Cornu}, \bibinfo{person}{Alan Arazi},
  \bibinfo{person}{Rylee Grace}, \bibinfo{person}{Simone Alessi},
  \bibinfo{person}{Mihir Manium}, \bibinfo{person}{Bernhard Schölkopf},
  \bibinfo{person}{Yann LeCun}, \bibinfo{person}{Madelon Hulsebos},
  \bibinfo{person}{Sauraj Gambhir}, \bibinfo{person}{Noah Hollmann}, {and}
  \bibinfo{person}{Frank Hutter}.} \bibinfo{year}{2026}\natexlab{}.
\newblock \bibinfo{title}{{TabPFN-3.5}: Technical Report}.
\newblock \bibinfo{howpublished}{arXiv preprint arXiv:2609.17895}.
\newblock
\showeprint[arxiv]{2609.17895}~[cs.LG]
\urldef\tempurl%
\url{https://arxiv.org/abs/2609.17895}
\showURL{%
\tempurl}


\bibitem[Kluger and Bates(2026)]%
        {kluger2026estimation}
\bibfield{author}{\bibinfo{person}{Dan~M. Kluger} {and}
  \bibinfo{person}{Stephen Bates}.} \bibinfo{year}{2026}\natexlab{}.
\newblock \bibinfo{title}{M-estimation under {Two-Phase} Multiwave Sampling
  with Applications to {Prediction-Powered} Inference}.
\newblock \bibinfo{howpublished}{arXiv preprint arXiv:2602.16933}.
\newblock
\showeprint[arxiv]{2602.16933}~[stat.ME]
\urldef\tempurl%
\url{https://arxiv.org/abs/2602.16933}
\showURL{%
\tempurl}


\bibitem[Lares et~al\mbox{.}(2025)]%
        {lares2025feature}
\bibfield{author}{\bibinfo{person}{Oscar Lares}, \bibinfo{person}{Hao Zhen},
  \bibinfo{person}{Jidong~J. Yang}, {and} \bibinfo{person}{{Smart Mobility and
  Infrastructure Lab, College of Engineering, University of Georgia, Athens, GA
  30602, USA}}.} \bibinfo{year}{2025}\natexlab{}.
\newblock \showarticletitle{Feature group tabular transformer: a novel approach
  to traffic crash modeling and causality analysis}.
\newblock \bibinfo{journal}{\emph{Applied Computing and Intelligence}}
  \bibinfo{volume}{5}, \bibinfo{number}{1} (\bibinfo{year}{2025}),
  \bibinfo{pages}{29--56}.
\newblock
\showeprint[arxiv]{2412.06825}
\href{https://doi.org/10.3934/aci.2025003}{doi:\nolinkurl{10.3934/aci.2025003}}


\bibitem[Majidi et~al\mbox{.}(2025)]%
        {majidi2025predicting}
\bibfield{author}{\bibinfo{person}{Mohammad~Zana Majidi},
  \bibinfo{person}{Sajjad Karimi}, \bibinfo{person}{Teng Wang},
  \bibinfo{person}{Robert Kluger}, {and} \bibinfo{person}{Reginald
  Souleyrette}.} \bibinfo{year}{2025}\natexlab{}.
\newblock \bibinfo{title}{Predicting person-level injury severity using crash
  narratives: A balanced approach with roadway classification and natural
  language process techniques}.
\newblock \bibinfo{howpublished}{arXiv preprint arXiv:2509.07845}.
\newblock
\showeprint[arxiv]{2509.07845}~[cs.LG]
\urldef\tempurl%
\url{https://arxiv.org/abs/2509.07845}
\showURL{%
\tempurl}


\bibitem[Marucci-Wellman et~al\mbox{.}(2017)]%
        {marucciwellman2017classifying}
\bibfield{author}{\bibinfo{person}{Helen~R. Marucci-Wellman},
  \bibinfo{person}{Helen~L. Corns}, {and} \bibinfo{person}{Mark~R. Lehto}.}
  \bibinfo{year}{2017}\natexlab{}.
\newblock \showarticletitle{Classifying injury narratives of large
  administrative databases for surveillance—A practical approach combining
  machine learning ensembles and human review}.
\newblock \bibinfo{journal}{\emph{Accident Analysis \& Prevention}}
  \bibinfo{volume}{98} (\bibinfo{year}{2017}), \bibinfo{pages}{359--371}.
\newblock
\href{https://doi.org/10.1016/j.aap.2016.10.014}{doi:\nolinkurl{10.1016/j.aap.2016.10.014}}


\bibitem[{National Center for Statistics and Analysis}(2025)]%
        {nhtsa2025distracted}
\bibfield{author}{\bibinfo{person}{{National Center for Statistics and
  Analysis}}.} \bibinfo{year}{2025}\natexlab{}.
\newblock \bibinfo{booktitle}{\emph{Distracted Driving in 2023}}.
\newblock \bibinfo{type}{Research Note} DOT HS 813 703.
  \bibinfo{institution}{National Highway Traffic Safety Administration}.
\newblock
\urldef\tempurl%
\url{https://crashstats.nhtsa.dot.gov/Api/Public/ViewPublication/813703}
\showURL{%
\tempurl}


\bibitem[{National Highway Traffic Safety Administration}(2017)]%
        {nhtsa2017mmucc}
\bibfield{author}{\bibinfo{person}{{National Highway Traffic Safety
  Administration}}.} \bibinfo{year}{2017}\natexlab{}.
\newblock \bibinfo{booktitle}{\emph{{MMUCC} Guideline: Model Minimum Uniform
  Crash Criteria, Fifth Edition (2017)}}.
\newblock \bibinfo{type}{{T}echnical {R}eport} DOT HS 812 433.
  \bibinfo{institution}{National Highway Traffic Safety Administration}.
\newblock
\urldef\tempurl%
\url{https://crashstats.nhtsa.dot.gov/Api/Public/ViewPublication/812433}
\showURL{%
\tempurl}


\bibitem[Neyman(1934)]%
        {neyman1934two}
\bibfield{author}{\bibinfo{person}{Jerzy Neyman}.}
  \bibinfo{year}{1934}\natexlab{}.
\newblock \showarticletitle{On the Two Different Aspects of the Representative
  Method: The Method of Stratified Sampling and the Method of Purposive
  Selection}.
\newblock \bibinfo{journal}{\emph{Journal of the Royal Statistical Society}}
  \bibinfo{volume}{97}, \bibinfo{number}{4} (\bibinfo{year}{1934}),
  \bibinfo{pages}{558}.
\newblock
\href{https://doi.org/10.2307/2342192}{doi:\nolinkurl{10.2307/2342192}}


\bibitem[Neyman(1938)]%
        {neyman1938contribution}
\bibfield{author}{\bibinfo{person}{J. Neyman}.}
  \bibinfo{year}{1938}\natexlab{}.
\newblock \showarticletitle{Contribution to the Theory of Sampling Human
  Populations}.
\newblock \bibinfo{journal}{\emph{J. Amer. Statist. Assoc.}}
  \bibinfo{volume}{33}, \bibinfo{number}{201} (\bibinfo{year}{1938}),
  \bibinfo{pages}{101--116}.
\newblock
\href{https://doi.org/10.1080/01621459.1938.10503378}{doi:\nolinkurl{10.1080/01621459.1938.10503378}}


\bibitem[Oliaee et~al\mbox{.}(2023)]%
        {oliaee2023using}
\bibfield{author}{\bibinfo{person}{Amir~Hossein Oliaee},
  \bibinfo{person}{Subasish Das}, \bibinfo{person}{Jinli Liu}, {and}
  \bibinfo{person}{M.~Ashifur Rahman}.} \bibinfo{year}{2023}\natexlab{}.
\newblock \showarticletitle{Using Bidirectional Encoder Representations from
  Transformers ({BERT}) to classify traffic crash severity types}.
\newblock \bibinfo{journal}{\emph{Natural Language Processing Journal}}
  \bibinfo{volume}{3} (\bibinfo{year}{2023}), \bibinfo{pages}{100007}.
\newblock
\href{https://doi.org/10.1016/j.nlp.2023.100007}{doi:\nolinkurl{10.1016/j.nlp.2023.100007}}


\bibitem[Platt(2000)]%
        {platt2000probabilities}
\bibfield{author}{\bibinfo{person}{John~C. Platt}.}
  \bibinfo{year}{2000}\natexlab{}.
\newblock \showarticletitle{Probabilities for {SV} Machines}.
\newblock In \bibinfo{booktitle}{\emph{Advances in Large-Margin Classifiers}}.
  \bibinfo{publisher}{The MIT Press}, \bibinfo{pages}{61--74}.
\newblock
\href{https://doi.org/10.7551/mitpress/1113.003.0008}{doi:\nolinkurl{10.7551/mitpress/1113.003.0008}}


\bibitem[Prokhorenkova et~al\mbox{.}(2018)]%
        {prokhorenkova2018catboost}
\bibfield{author}{\bibinfo{person}{Liudmila Prokhorenkova},
  \bibinfo{person}{Gleb Gusev}, \bibinfo{person}{Aleksandr Vorobev},
  \bibinfo{person}{Anna~Veronika Dorogush}, {and} \bibinfo{person}{Andrey
  Gulin}.} \bibinfo{year}{2018}\natexlab{}.
\newblock \showarticletitle{{CatBoost}: unbiased boosting with categorical
  features}. In \bibinfo{booktitle}{\emph{Advances in Neural Information
  Processing Systems}}, Vol.~\bibinfo{volume}{31}.
\newblock
\showeprint[arxiv]{1706.09516}
\urldef\tempurl%
\url{https://papers.nips.cc/paper_files/paper/2018/hash/14491b756b3a51daac41c24863285549-Abstract.html}
\showURL{%
\tempurl}


\bibitem[Purucker et~al\mbox{.}(2026)]%
        {purucker2026beyond}
\bibfield{author}{\bibinfo{person}{Lennart Purucker}, \bibinfo{person}{Andrej
  Tschalzev}, \bibinfo{person}{Nick Erickson}, \bibinfo{person}{Gioia Blayer},
  \bibinfo{person}{David Holzmüller}, \bibinfo{person}{Alan Arazi},
  \bibinfo{person}{Alexander Pfefferle}, \bibinfo{person}{Mustafa Tajjar},
  \bibinfo{person}{Gaël Varoquaux}, {and} \bibinfo{person}{Frank Hutter}.}
  \bibinfo{year}{2026}\natexlab{}.
\newblock \bibinfo{title}{Beyond {IID}: How General Are Tabular Foundation
  Models, Really?}
\newblock \bibinfo{howpublished}{arXiv preprint arXiv:2606.30410}.
\newblock
\showeprint[arxiv]{2606.30410}~[cs.LG]
\urldef\tempurl%
\url{https://arxiv.org/abs/2606.30410}
\showURL{%
\tempurl}


\bibitem[Qu et~al\mbox{.}(2026a)]%
        {qu2026tabiclv2}
\bibfield{author}{\bibinfo{person}{Jingang Qu}, \bibinfo{person}{David
  Holzmüller}, \bibinfo{person}{Gaël Varoquaux}, {and}
  \bibinfo{person}{Marine Le~Morvan}.} \bibinfo{year}{2026}\natexlab{a}.
\newblock \showarticletitle{{TabICLv2}: A Better, Faster, Scalable, and Open
  Tabular Foundation Model}. In \bibinfo{booktitle}{\emph{International
  Conference on Machine Learning (ICML 2026)}}.
\newblock
\showeprint[arxiv]{2602.11139}
\urldef\tempurl%
\url{https://openreview.net/forum?id=SxsyLjIfWB}
\showURL{%
\tempurl}


\bibitem[Qu et~al\mbox{.}(2026b)]%
        {qu2026nvidia}
\bibfield{author}{\bibinfo{person}{Jingang Qu}, \bibinfo{person}{Valter
  Hudovernik}, \bibinfo{person}{Martin Jurkovic}, \bibinfo{person}{Cedric
  Lorenz}, \bibinfo{person}{Akihiro Nitta}, \bibinfo{person}{Dmitry Gordeev},
  \bibinfo{person}{Federico Lopez}, \bibinfo{person}{Ramona Bendias},
  \bibinfo{person}{Gilberto Titericz~Jr}, \bibinfo{person}{Aleksandar~S.
  Sokolovski}, \bibinfo{person}{Isabel Hulseman}, \bibinfo{person}{Jure
  Leskovec}, {and} \bibinfo{person}{Matthias Fey}.}
  \bibinfo{year}{2026}\natexlab{b}.
\newblock \bibinfo{title}{{NVIDIA} {Kumo} {Tabular} Sets a New
  Accuracy-Efficiency Frontier for Tabular Prediction}.
\newblock \bibinfo{howpublished}{Hugging Face blog, published 29 September
  2026}.
\newblock
\urldef\tempurl%
\url{https://huggingface.co/blog/nvidia/kumo-tabular}
\showURL{%
\tempurl}


\bibitem[Rafe et~al\mbox{.}(2024)]%
        {rafe2024comparative}
\bibfield{author}{\bibinfo{person}{Amir Rafe}, \bibinfo{person}{Mohammad~Ali
  Arman}, {and} \bibinfo{person}{Patrick~A. Singleton}.}
  \bibinfo{year}{2024}\natexlab{}.
\newblock \showarticletitle{A Comparative Study Using Generalized Ordered
  Probit, Stacking Ensemble, and {TabNet}: Application to Determinants of
  Pedestrian Crash Severity}.
\newblock \bibinfo{journal}{\emph{Data Science for Transportation}}
  \bibinfo{volume}{6}, \bibinfo{number}{2} (\bibinfo{year}{2024}),
  \bibinfo{pages}{13}.
\newblock
\href{https://doi.org/10.1007/s42421-024-00098-x}{doi:\nolinkurl{10.1007/s42421-024-00098-x}}


\bibitem[Rafe and Das(2026a)]%
        {rafe2026calibrated}
\bibfield{author}{\bibinfo{person}{Amir Rafe} {and} \bibinfo{person}{Subasish
  Das}.} \bibinfo{year}{2026}\natexlab{a}.
\newblock \bibinfo{title}{Calibrated Decisions at Scale: Converting Police
  Crash Narratives into Probabilistic Crash Variables with a System One Model
  (Jev)}.
\newblock \bibinfo{howpublished}{arXiv preprint arXiv:2609.24052}.
\newblock
\showeprint[arxiv]{2609.24052}~[cs.CL]
\urldef\tempurl%
\url{https://arxiv.org/abs/2609.24052}
\showURL{%
\tempurl}


\bibitem[Rafe and Das(2026b)]%
        {rafe2026counting}
\bibfield{author}{\bibinfo{person}{Amir Rafe} {and} \bibinfo{person}{Subasish
  Das}.} \bibinfo{year}{2026}\natexlab{b}.
\newblock \bibinfo{title}{Counting the Uncounted: Population-Level Surveillance
  of Documented Pregnancy and Fetal Harm in Police Crash Narratives with a
  System One Model ({Jev})}.
\newblock \bibinfo{howpublished}{arXiv preprint arXiv:2610.00213}.
\newblock
\showeprint[arxiv]{2610.00213}~[stat.AP]
\urldef\tempurl%
\url{https://arxiv.org/abs/2610.00213}
\showURL{%
\tempurl}


\bibitem[Rolison et~al\mbox{.}(2018)]%
        {rolison2018what}
\bibfield{author}{\bibinfo{person}{Jonathan~J. Rolison},
  \bibinfo{person}{Shirley Regev}, \bibinfo{person}{Salissou Moutari}, {and}
  \bibinfo{person}{Aidan Feeney}.} \bibinfo{year}{2018}\natexlab{}.
\newblock \showarticletitle{What are the factors that contribute to road
  accidents? An assessment of law enforcement views, ordinary drivers’
  opinions, and road accident records}.
\newblock \bibinfo{journal}{\emph{Accident Analysis \& Prevention}}
  \bibinfo{volume}{115} (\bibinfo{year}{2018}), \bibinfo{pages}{11--24}.
\newblock
\href{https://doi.org/10.1016/j.aap.2018.02.025}{doi:\nolinkurl{10.1016/j.aap.2018.02.025}}


\bibitem[Saroj et~al\mbox{.}(2026)]%
        {saroj2026adapting}
\bibfield{author}{\bibinfo{person}{Abhilasha Saroj}, \bibinfo{person}{Pranav
  Govindu}, \bibinfo{person}{Bharat Sharma}, {and} \bibinfo{person}{Usman
  Ahmed}.} \bibinfo{year}{2026}\natexlab{}.
\newblock \bibinfo{title}{Adapting a Large Language Model {Crash-Severity}
  Pipeline to Tennessee: Performance Across Sampling Strategies}.
\newblock \bibinfo{howpublished}{arXiv preprint arXiv:2609.29793}.
\newblock
\showeprint[arxiv]{2609.29793}~[cs.CE]
\urldef\tempurl%
\url{https://arxiv.org/abs/2609.29793}
\showURL{%
\tempurl}


\bibitem[Schiff and Cummings(2004)]%
        {schiff2004comparison}
\bibfield{author}{\bibinfo{person}{Melissa~A. Schiff} {and}
  \bibinfo{person}{Peter Cummings}.} \bibinfo{year}{2004}\natexlab{}.
\newblock \showarticletitle{Comparison of reporting of seat belt use by police
  and crash investigators: variation in agreement by injury severity}.
\newblock \bibinfo{journal}{\emph{Accident Analysis \& Prevention}}
  \bibinfo{volume}{36}, \bibinfo{number}{6} (\bibinfo{year}{2004}),
  \bibinfo{pages}{961--965}.
\newblock
\href{https://doi.org/10.1016/j.aap.2003.09.007}{doi:\nolinkurl{10.1016/j.aap.2003.09.007}}


\bibitem[Shi et~al\mbox{.}(2026)]%
        {shi2026multi}
\bibfield{author}{\bibinfo{person}{Yan Shi}, \bibinfo{person}{Cheng Liu},
  \bibinfo{person}{Aodi Yu}, \bibinfo{person}{Zhenzhou Lu},
  \bibinfo{person}{Said Elias}, \bibinfo{person}{Kai Cheng},
  \bibinfo{person}{Jiaqing Kou}, \bibinfo{person}{Xin Chen},
  \bibinfo{person}{Yu Liu}, \bibinfo{person}{Hong-Zhong Huang}, {and}
  \bibinfo{person}{Michael Beer}.} \bibinfo{year}{2026}\natexlab{}.
\newblock \showarticletitle{A multi-fidelity tabular prior-data fitted network
  model for accurate prediction and uncertainty quantification}.
\newblock \bibinfo{journal}{\emph{Nature Communications}} \bibinfo{volume}{17},
  \bibinfo{number}{1} (\bibinfo{year}{2026}), \bibinfo{pages}{8308}.
\newblock
\href{https://doi.org/10.1038/s41467-026-75163-w}{doi:\nolinkurl{10.1038/s41467-026-75163-w}}


\bibitem[Sjögren et~al\mbox{.}(1997)]%
        {sjogren1997comparison}
\bibfield{author}{\bibinfo{person}{Harmeet Sjögren}, \bibinfo{person}{Ulf
  Björnstig}, {and} \bibinfo{person}{Anders Eriksson}.}
  \bibinfo{year}{1997}\natexlab{}.
\newblock \showarticletitle{Comparison between blood analysis and police
  assessment of drug and alcohol use by injured drivers}.
\newblock \bibinfo{journal}{\emph{Scandinavian Journal of Social Medicine}}
  \bibinfo{volume}{25}, \bibinfo{number}{3} (\bibinfo{year}{1997}),
  \bibinfo{pages}{217--223}.
\newblock
\href{https://doi.org/10.1177/140349489702500312}{doi:\nolinkurl{10.1177/140349489702500312}}


\bibitem[Somvanshi et~al\mbox{.}(2025a)]%
        {somvanshi2025tabular}
\bibfield{author}{\bibinfo{person}{Shriyank Somvanshi}, \bibinfo{person}{Pavan
  Hebli}, \bibinfo{person}{Gaurab Chhetri}, {and} \bibinfo{person}{Subasish
  Das}.} \bibinfo{year}{2025}\natexlab{a}.
\newblock \showarticletitle{Tabular Data with Class Imbalance: Predicting
  Electric Vehicle Crash Severity with Pretrained Transformers ({TabPFN}) and
  {Mamba-Based} Models}. In \bibinfo{booktitle}{\emph{2025 International
  Conference on Machine Learning and Applications (ICMLA)}}.
  \bibinfo{publisher}{IEEE}, \bibinfo{pages}{1460--1465}.
\newblock
\showeprint[arxiv]{2509.11449}
\href{https://doi.org/10.1109/icmla66185.2025.00222}{doi:\nolinkurl{10.1109/icmla66185.2025.00222}}


\bibitem[Somvanshi et~al\mbox{.}(2025b)]%
        {somvanshi2025applying}
\bibfield{author}{\bibinfo{person}{Shriyank Somvanshi},
  \bibinfo{person}{Anannya~Ghosh Tusti}, \bibinfo{person}{Mahmuda~Sultana
  Mimi}, \bibinfo{person}{Md~Monzurul Islam}, \bibinfo{person}{Sazzad
  Bin~Bashar Polock}, \bibinfo{person}{Anandi Dutta}, {and}
  \bibinfo{person}{Subasish Das}.} \bibinfo{year}{2025}\natexlab{b}.
\newblock \bibinfo{title}{Applying {MambaAttention}, {TabPFN}, and
  {TabTransformers} to Classify {SAE} Automation Levels in Crashes}.
\newblock \bibinfo{howpublished}{arXiv preprint arXiv:2506.03160}.
\newblock
\showeprint[arxiv]{2506.03160}~[cs.LG]
\urldef\tempurl%
\url{https://arxiv.org/abs/2506.03160}
\showURL{%
\tempurl}


\bibitem[Särndal et~al\mbox{.}(1992)]%
        {sarndal1992model}
\bibfield{author}{\bibinfo{person}{Carl-Erik Särndal}, \bibinfo{person}{Bengt
  Swensson}, {and} \bibinfo{person}{Jan Wretman}.}
  \bibinfo{year}{1992}\natexlab{}.
\newblock \bibinfo{booktitle}{\emph{Model Assisted Survey Sampling}}.
\newblock \bibinfo{publisher}{Springer New York}.
\newblock
\href{https://doi.org/10.1007/978-1-4612-4378-6}{doi:\nolinkurl{10.1007/978-1-4612-4378-6}}


\bibitem[{Texas Department of Transportation}(2025)]%
        {txdot2025instructions}
\bibfield{author}{\bibinfo{person}{{Texas Department of Transportation}}.}
  \bibinfo{year}{2025}\natexlab{}.
\newblock \bibinfo{booktitle}{\emph{Instructions to Police for Reporting
  Crashes ({CR-100}), Version 29.0}}.
\newblock Texas Department of Transportation, Traffic Safety Division.
\newblock
\urldef\tempurl%
\url{https://www.txdot.gov/content/dam/docs/division/trf/crash-records/cr100-v29.pdf}
\showURL{%
\tempurl}
\newblock
\shownote{2025 edition, revised 4 June 2025}.


\bibitem[{TypeSafe}(2026a)]%
        {typesafe2026launch}
\bibfield{author}{\bibinfo{person}{{TypeSafe}}.}
  \bibinfo{year}{2026}\natexlab{a}.
\newblock \bibinfo{title}{Introducing System One Models and Jev}.
\newblock \bibinfo{howpublished}{Vendor blog post}.
\newblock
\urldef\tempurl%
\url{https://typesafe.ai/blog/introducing-system-one-models-and-jev}
\showURL{%
\tempurl}


\bibitem[{TypeSafe}(2026b)]%
        {typesafe2026systemone}
\bibfield{author}{\bibinfo{person}{{TypeSafe}}.}
  \bibinfo{year}{2026}\natexlab{b}.
\newblock \bibinfo{title}{System One}.
\newblock \bibinfo{howpublished}{Vendor documentation}.
\newblock
\urldef\tempurl%
\url{https://docs.typesafe.ai/concepts/system-one}
\showURL{%
\tempurl}


\bibitem[Zhen and Yang(2025)]%
        {zhen2025crashsage}
\bibfield{author}{\bibinfo{person}{Hao Zhen} {and} \bibinfo{person}{Jidong~J.
  Yang}.} \bibinfo{year}{2025}\natexlab{}.
\newblock \showarticletitle{{CrashSage}: A large language model-centered
  framework for contextual and interpretable traffic crash analysis}.
\newblock \bibinfo{journal}{\emph{Artificial Intelligence for Transportation}}
  \bibinfo{volume}{3-4} (\bibinfo{year}{2025}), \bibinfo{pages}{100030}.
\newblock
\href{https://doi.org/10.1016/j.ait.2025.100030}{doi:\nolinkurl{10.1016/j.ait.2025.100030}}


\bibitem[Zrnic and Candes(2024)]%
        {zrnic2024active}
\bibfield{author}{\bibinfo{person}{Tijana Zrnic} {and}
  \bibinfo{person}{Emmanuel Candes}.} \bibinfo{year}{2024}\natexlab{}.
\newblock \showarticletitle{Active Statistical Inference}. In
  \bibinfo{booktitle}{\emph{Proceedings of the 41st International Conference on
  Machine Learning}} \emph{(\bibinfo{series}{Proceedings of Machine Learning
  Research}, Vol.~\bibinfo{volume}{235})}. \bibinfo{publisher}{PMLR},
  \bibinfo{pages}{62993--63010}.
\newblock
\showeprint[arxiv]{2403.03208}
\urldef\tempurl%
\url{https://proceedings.mlr.press/v235/zrnic24a.html}
\showURL{%
\tempurl}


\bibitem[Zrnic and Candès(2024)]%
        {zrnic2024cross}
\bibfield{author}{\bibinfo{person}{Tijana Zrnic} {and}
  \bibinfo{person}{Emmanuel~J. Candès}.} \bibinfo{year}{2024}\natexlab{}.
\newblock \showarticletitle{Cross-prediction-powered inference}.
\newblock \bibinfo{journal}{\emph{Proceedings of the National Academy of
  Sciences}} \bibinfo{volume}{121}, \bibinfo{number}{15}
  (\bibinfo{year}{2024}), \bibinfo{pages}{e2322083121}.
\newblock
\href{https://doi.org/10.1073/pnas.2322083121}{doi:\nolinkurl{10.1073/pnas.2322083121}}


\end{thebibliography}

\appendix
\counterwithin{table}{section}
\counterwithin{figure}{section}
\counterwithin{equation}{section}
\section{The coded feature set}\label{app:features}

The table reader sees \NFeatures{} features of the coded record, all taken from the Texas crash extract. Thirty-two are crash fields, such as the hour, road type, surface, light, weather, speed limit, manner of collision, and the response time of the investigating officer. Eighteen are unit fields aggregated over the distinct units of a crash, such as vehicle body types and counts, and twelve are families of coded contributing factors marked when any unit carries them. Twenty-six are person fields aggregated over the distinct persons, such as age, injury severity, restraint use, and test results, and nine are binary indicators of the factors derived from the coded fields, the other seven of sixteen such indicators duplicating a column above. Categorical fields keep their coded labels, and the labels Unknown, Not Reported, and No Data become missing. Narrative text, identifiers, coordinates, license numbers, dates of birth, ZIP codes, and ethnicity are never read. The released code lists every feature with its source field and flattening rule.

\section{Estimator, recalibration, and their uncertainty}\label{app:estimator}\label{app:recal}

\begin{table}[h]
\caption{Notation used in the estimator and the evaluation.}\label{tab:notation}
\tablestyle
\begin{tabular}{P{0.18}P{0.82}}
\toprule
\headrow \hd{Symbol} & \hd{Meaning} \\
\midrule
$U$, $N$ & Geolocated crashes with a narrative of more than forty characters, and their number \\
$x_i$ & The coded record of crash $i$ (\NFeatures{} features) \\
$v$ & A factor, such as alcohol or unbelted occupants \\
$c_{vi}$ & The coded-field indicator for $v$, defined for the nine factors with a coded counterpart \\
$p_{vi}$ & The narrative reader's probability that the narrative states $v$ \\
$\tilde q_{vi}$ & The recalibrated probability $g_v(p_{vi}, c_{vi})$, the phase-1 measurement \\
$y_{vi}$ & The human judgment that the narrative states $v$ (phase 2) \\
$\hat f_{vi}$ & The table reader's probability for crash $i$, known for every crash (phase 0) \\
$I^{(1)}_i$, $\pi^{(1)}_i$ & Wave-1 indicator and inclusion probability, $\pi^{(1)}_i = n_1 / N$ \\
$I^{(2)}_i$, $\pi^{(2)}_i$ & Wave-2 indicator and recorded per-unit inclusion probability \\
$\hat f^{\,d}_{vi}$ & The table reader's probability of confirmed discordance, used for the second wave and the re-read list \\
$W_i$ & The multiwave inverse probability weight of Eq.~\eqref{eq:weights} \\
$\delta_i$ & A domain indicator, such as injury-or-fatal crashes of one year \\
$T_{v\delta}$ & The estimand, the number of crashes in domain $\delta$ whose narrative states $v$ \\
$\Omega$, $\lambda$ & The coefficients of the table term in the multiwave and the sequential estimator \\
$\eta$ & The coded-field shift of the recalibration map \\
\bottomrule
\end{tabular}
\tablenote{Indices $i$ run over crashes and $v$ over factors. A tilde marks a recalibrated quantity and a hat marks a table-reader output.}
\end{table}

This appendix states why the estimator of Eq.~\eqref{eq:mpd} is consistent for the recalibrated total, how its variance and the recalibration variance are estimated, and how the recalibration map was chosen. The argument follows Kluger and Bates~\cite{kluger2026estimation}, specialized to a domain mean. Throughout, $\theta = \mathbb{E}[\delta\tilde q]$ and $\gamma = \mathbb{E}[\delta \hat f]$ over $U$, so that $T_{v\delta} = N\theta$.

Each wave term of Eq.~\eqref{eq:weights} has conditional expectation one given the earlier waves. For wave 1, $\mathbb{E}[I^{(1)}_i / \pi^{(1)}_i] = 1$. For wave 2, conditional on wave 1, $\mathbb{E}[(1 - I^{(1)}_i)/(1 - \pi^{(1)}_i) \cdot I^{(2)}_i / \pi^{(2)}_i] = (1 - I^{(1)}_i)/(1 - \pi^{(1)}_i)$, whose expectation over wave 1 is again one. Hence $\mathbb{E}[W_i] = c_1 + c_2 = 1$ for every unit, the ratio estimators $\hat\theta_{\mathrm{II}}$ and $\hat\gamma_{\mathrm{II}}$ are consistent for $\theta$ and $\gamma$, and $\bar\gamma_{\mathrm{I}}$ equals $\gamma$ exactly, so for any fixed $\Omega$ the estimator converges to $\Omega\gamma + \theta - \Omega\gamma = \theta$. Cross-fitting makes each wave-1 prediction independent of that crash's own measurement, and the five fold models differ little, which is the stability condition of the cross-prediction guarantees~\cite{zrnic2024cross}. With $\Omega$ tuned on the same sample and restricted to the unit interval, consistency still holds, because it holds for every fixed value in the interval and the tuned value converges.

The variance follows from the asymptotic linear expansion of Kluger and Bates. With the squared-error loss the gradient is $-2(x - \theta)$ and the Hessian is two, and their variance estimator takes the following form.
\begin{equation}\label{eq:mpdvar}
\hat\Sigma = \frac{\hat\Sigma_{11}}{\hat H_\theta^2} + \Omega^2 \frac{\hat\Sigma_{22} - \hat\Sigma_{33}}{4} + 2\Omega\,\frac{\hat\Sigma_{13} - \hat\Sigma_{12}}{2\hat H_\theta}.
\end{equation}
Here $\hat\Sigma_{11}$, $\hat\Sigma_{12}$, and $\hat\Sigma_{22}$ are weighted second moments of the loss gradients with weights $W_i^2$, $\hat\Sigma_{13}$ uses weights $W_i$, $\hat\Sigma_{33}$ is the unweighted second moment of $\delta\hat f$ over $U$, and $\hat H_\theta = 2\sum_i W_i / N$. The design variance of $\hat T_{v\delta}$ is $N\hat\Sigma$, without a finite-population correction for wave 1, which makes it slightly conservative. The variance-minimizing $\Omega$ is $-(\hat\Sigma_{13} - \hat\Sigma_{12}) / (\hat H_\theta (\hat\Sigma_{22} - \hat\Sigma_{33})/2)$, restricted to the unit interval.

The recalibration map was chosen by held-out Brier score over fifty narrative half-splits. Because the target is a sum, a candidate was eligible only if its held-out calibration in the large, the weighted mean of $\tilde q_{vj} - y_{vj}$, was consistent with zero at the 95 percent level, judged by its narrative-clustered standard error. Among eligible candidates the simplest whose Brier score exceeded the best by no more than one narrative-clustered standard error of the paired difference was kept, and when none was eligible, the one with the smallest standardized bias was kept, which happened only for unbelted occupants. The identity was ineligible for every factor except vehicle defects, because the narrative reader's probabilities overstate the human rate. Table~\ref{tab:recal} gives the chosen maps. Among the recalibrated maps fitted on all the judgments, the estimates move by at most \MapSpreadMax{} percent for every factor except vehicle defects, whose isotonic map gives a total \MapSpreadDefect{} percent lower. Letting each coded factor's shift differ from the shared $\eta$ moves its estimate by at most \ShiftMaxChange{} percent, for \ShiftMaxFactor{}, and leaves all nine estimates inside their intervals.

\begin{table}[t]
\caption{Recalibration maps and the confirmation of Jev flags by coded-field status.}\label{tab:recal}
\tablestyle
\begin{tabular}{P{0.17}P{0.17}R{0.10}R{0.09}R{0.09}R{0.19}R{0.19}}
\toprule
\headrow \hd{} & \hd{} & \hd{} & \multicolumn{2}{c}{\hdtop{Held-out Brier}} & \multicolumn{2}{c}{\hdtop{Flags confirmed}} \\
\headrow \hdsub{Factor} & \hdsub{Map} & \hdsub{Positives} & \hdsub{Raw} & \hdsub{Chosen} & \hdsub{Field true} & \hdsub{Field false} \\
\midrule
Aggression & Platt & 31 & 3.69 & 1.75 & \cellnote{none} & \cellnote{none} \\
Alcohol & Platt, coded shift & 25 & 4.56 & 2.71 & \cellbar[humanslate]{1.00}{100\% (17)} & \cellbar[humanslate]{0.71}{71\% (13)} \\
Animal & Platt & 23 & 0.57 & 0.35 & \cellbar[humanslate]{1.00}{100\% (22)} & \cellnote{none} \\
Drugs$^{\dagger}$ & Isotonic & 8 & 1.20 & 0.00 & \cellbar[humanslate]{1.00}{100\% (7)} & \cellbar[humanslate]{0.52}{52\% (2)} \\
Fatigue & Platt & 27 & 0.84 & 0.42 & \cellbar[humanslate]{0.96}{96\% (18)} & \cellbar[humanslate]{0.89}{89\% (12)} \\
Hit-and-run & Platt & 92 & 7.44 & 5.84 & \cellnote{none} & \cellnote{none} \\
Hydroplane & Platt, coded shift & 31 & 1.28 & 0.98 & \cellbar[humanslate]{1.00}{100\% (20)} & \cellbar[humanslate]{0.88}{88\% (11)} \\
Intentional$^{\dagger}$ & Platt & 5 & 3.77 & 0.62 & \cellnote{none} & \cellnote{none} \\
Medical episode & Platt & 24 & 2.82 & 1.44 & \cellbar[humanslate]{1.00}{100\% (12)} & \cellbar[humanslate]{0.57}{57\% (18)} \\
Phone use$^{\dagger}$ & Platt & 15 & 0.88 & 0.29 & \cellbar[humanslate]{0.96}{96\% (8)} & \cellbar[humanslate]{0.90}{90\% (9)} \\
Transported & Platt & 35 & 2.01 & 0.81 & \cellnote{none} & \cellnote{none} \\
Unbelted$^{\dagger}$ & Isotonic & 5 & 3.59 & 0.14 & \cellbar[humanslate]{1.00}{100\% (3)} & \cellbar[humanslate]{0.15}{15\% (13)} \\
Vehicle defect & Identity & 32 & 17.26 & 17.26 & \cellnote{none} & \cellnote{none} \\
Witness cited & Platt & 41 & 11.86 & 11.00 & \cellnote{none} & \cellnote{none} \\
Wrong-way & Platt & 21 & 9.57 & 5.92 & \cellbar[humanslate]{1.00}{100\% (8)} & \cellbar[humanslate]{0.26}{26\% (27)} \\
\bottomrule
\end{tabular}
\tablenote{Brier scores are multiplied by one thousand and are held out over fifty narrative half-splits. The coded-field shift is estimated jointly over the nine factors with a coded field, $\eta = \GammaShift{}$ (95 percent interval \GammaLo{} to \GammaHi{}), and it enters the chosen map of the factors marked Platt, coded shift. Flags confirmed is the calibration-weighted share of Jev flags at $p \ge 0.5$ that the human tier confirmed, split by whether the coded field also records the factor, drawn as a bar whose full track is 100 percent, with the number of judged flags in parentheses. The shares rest on few flags and carry wide intervals, which the released table gives. $^{\dagger}$Pooled map.}
\end{table}

The recalibration variance comes from \BRecal{} narrative cluster resamples of the human judgments within their draw strata. Each resample reapplies the twenty-positive pooling rule, reselects the map by the same rule over the same fifty half-splits, refits it, and recomputes $\hat T_{v\delta}$ with $\Omega$ held at its point value. Each bound of the interval adds, in quadrature, the normal design half-width and the distance from $\hat T_{v\delta}$ to the matching end of the central 95 percent range of the refit totals. For symmetric refits this is $\hat T_{v\delta} \pm 1.96$ times the square root of the summed variances, and for skewed refits it keeps each bound within the range that the refits support.

The retrospective comparison of human-tier designs asks what a human tier stratified on the disagreement between the two readers, $|\hat f_{vi} - \tilde q_{vi}|$, would have bought against the stratification on the narrative reader's probability that drew the phase-2 tier. It treats the human-judged narratives as a pseudo-population and compares the Neyman variance of the mean held-out residual $e = y - \tilde q$, the quantity that drives the recalibration error of a total to first order, under the two designs of the same size. For strata $h$ with weighted shares $W_h$ and weighted standard deviations $S_h$ and $n$ judgments,
\begin{equation}\label{eq:neyman}
V = \frac{1}{n}\Big(\sum_h W_h S_h\Big)^2,
\end{equation}
and the ratio of $V$ under the two stratifications, with a narrative cluster bootstrap interval, is reported.

\section{Sampling designs of the second wave and the second human tier}\label{app:wave2}\label{app:round2}

The second-wave frame was the \WaveTwoFrame{} narratives of $U$ outside wave 1. The constant $\kappa$ of Eq.~\eqref{eq:pi2} was found by bisection so that $\sum_i \min(1, \kappa s_i)$ equals \nTwoExpScore{}, and $c_0$ was set to \nTwoExpControl{} divided by the frame size. Each unit received its own uniform variate from a generator with a fixed seed and was drawn when the variate fell below $\pi^{(2)}_i$. The probabilities ranged from \PiTwoMin{} to \PiTwoMax{}, so none reached the cap of \PiCap{}. Of the \nTwoDraws{} draws, \nTwoReuse{} reused existing reads from enriched samples, and the other \nTwoNew{} were read for this paper with every call recording the model identifier.

The re-read list is defined on the unread narratives, which are known only after wave 1, so its totals use the sequential estimator. For a domain $\delta$ inside the unread frame $F$,
\begin{equation}\label{eq:seq}
\hat T^{\mathrm{seq}}_{v\delta} = \sum_{i \in F} \frac{I^{(2)}_i}{\pi^{(2)}_i}\, \delta_i \big(\tilde q_{vi} - \lambda \hat f_{vi}\big) + \lambda \sum_{i \in F} \delta_i \hat f_{vi},
\end{equation}
which is unbiased conditional on wave 1 for any fixed $\lambda$ because every unit of $F$ has $\pi^{(2)}_i > 0$. Its variance is the Poisson-design variance $\sum_{i \in F} I^{(2)}_i (1 - \pi^{(2)}_i)\, \delta_i (\tilde q_{vi} - \lambda \hat f_{vi})^2 / (\pi^{(2)}_i)^2$, and $\lambda$ is the weighted regression slope of $\delta\tilde q$ on $\delta\hat f$ among the measured units, restricted to the unit interval. In a simulation, one synthetic population of a hundred thousand units was held fixed and the design redrawn \SimDesigns{} times. The 95 percent intervals covered the truth in \CovWaveOne{} of designs for wave 1, \CovSeq{} for the sequential estimator, and \CovMPD{} for the multiwave estimator, with a Monte Carlo standard error of about one percentage point, and the second wave narrowed the multiwave interval by \SimNarrowing{} percent.

The second human tier was drawn from the \RtwoFrame{} wave-1 narratives outside the phase-2 tier, with Poisson inclusion probabilities proportional to $\sqrt{\sum_v \mathbf{1}[p_{vi} \ge \ActiveThreshold{}]\, \tilde q_{vi}(1 - \tilde q_{vi}) / T_v^2}$, the compromise Neyman allocation of the correction over the fifteen factors. A floor keeps every probability positive, and the largest probability is \RtwoPiMax{}. Of the \RtwoDrawn{} narratives drawn, the identifier screen, strengthened for this tier by rules for addresses, phone numbers, plates, and written dates, withheld \RtwoWithheld{}, or \RtwoWithheldPct{} percent, which the weights treat as nonresponse within quintile classes of the inclusion probability. Of the cleared narratives, \RtwoCoded{} were kept in a recorded random order. Each narrative was assigned to two of the three coders, and every narrative that the third coder read was also read by a specialist, so a specialist read \RtwoNarr{} narratives with \RtwoJudg{} usable judgments. The narratives that only the two specialists shared were coded in the recorded random order to \RtwoCompletionCC{} percent, a known share that enters the weights.

The third coder agreed with the specialists at a Cohen's $\kappa$ between \KappaNonspecLo{} and \KappaNonspecHi{}, against \KappaSpec{} between the specialists on \NSpecShared{} shared judgments, so the specialist's label is the reference. In the phase-2 tier the reference is the agreement of the coders, who agreed on \RoneAgree{} percent of pairs, with the \RoneDisputed{} disputed pairs left out. Two sensitivity analyses apply the phase-2 tier's agreement rule and the third coder's labels to the second tier.

\section{Further conditions of the table-reader evaluation}\label{app:reader}

Table~\ref{tab:readertargets} gives the AUC of every reader on every target at the shared context of \BenchCtx{} rows, with the paired comparisons against Kumo Tabular small that the main text counts. Calibration is measured against the narrative reader's flag, the label the table reader is trained on. Kumo Tabular small, TabPFN 3.5, and CatBoost have mean calibration errors between \EcePFNNew{} and \EceCatBoost{} and mean slopes of \SlopeKumo{}, \SlopePFNNew{}, and \SlopeCatBoost{}, while TabICLv2 is overconfident with a slope of \SlopeTabICL{}. Context matters more than model size. The mean flag AUC of Kumo Tabular small rises from \AucFlagKumoThousand{} at the smallest context to \AucFlagKumo{} at \BenchCtx{} rows, while at ten thousand rows the small, medium, and large models reach \AucFlagKumoTen{}, \AucFlagKumoMediumTen{}, and \AucFlagKumoLargeTen{}, and the large model needs \GpuKumoLargeTen{} GB of memory.

\begin{table}[h]
\caption{AUC of each table reader on each target.}\label{tab:readertargets}
\tablestyle
\begin{tabular}{P{0.2}R{0.1}R{0.12}R{0.11}R{0.11}R{0.11}R{0.15}}
\toprule
\headrow \hdsub{Target} & \hdsub{Kumo small} & \hdsub{TabPFN 3.5} & \hdsub{TabPFN 3} & \hdsub{TabICLv2} & \hdsub{CatBoost} & \hdsub{CatBoost pool} \\
\midrule
\grouprow\multicolumn{7}{l}{\grouplabel{Narrative flags}}\\
Alcohol & 0.963 & \best{0.964} & 0.960$^{*}$ & 0.959$^{*}$ & 0.957$^{*}$ & 0.958$^{*}$ \\
Drugs & 0.900 & \best{0.910} & 0.905 & 0.903 & 0.898 & 0.905 \\
Hit-and-run & 0.932 & \best{0.935}$^{*}$ & 0.930$^{*}$ & 0.928$^{*}$ & 0.927$^{*}$ & 0.931 \\
Medical episode & 0.869 & \best{0.879}$^{*}$ & 0.871 & 0.869 & 0.858 & 0.864 \\
Phone use & 0.891 & \best{0.892} & 0.888 & 0.888 & 0.884 & 0.883 \\
Unbelted & \best{0.768} & 0.767 & 0.731$^{*}$ & 0.682$^{*}$ & 0.729$^{*}$ & 0.737$^{*}$ \\
Witness cited & 0.783 & \best{0.797}$^{*}$ & 0.783 & 0.770$^{*}$ & 0.786 & 0.791$^{*}$ \\
Wrong-way & 0.816 & \best{0.828}$^{*}$ & 0.807 & 0.803$^{*}$ & 0.791$^{*}$ & 0.797$^{*}$ \\
\grouprow\multicolumn{7}{l}{\grouplabel{Discordance, coded field false}}\\
Phone use & 0.834 & \best{0.837} & 0.829 & 0.826 & 0.818 & 0.828 \\
Unbelted & 0.701 & \best{0.709} & 0.674$^{*}$ & 0.578$^{*}$ & 0.645$^{*}$ & 0.651$^{*}$ \\
Wrong-way & 0.743 & \best{0.769}$^{*}$ & 0.734 & 0.722$^{*}$ & 0.710$^{*}$ & 0.731 \\
\bottomrule
\end{tabular}
\tablenote{All readers use \BenchCtx{} context rows, except CatBoost pool, which was trained on the whole pool. Bold marks the highest AUC in each row at printed precision. An asterisk marks a paired bootstrap difference from Kumo Tabular small whose 95 percent interval excludes zero.}
\end{table} The soft-label condition trained Kumo Tabular small as a regression on the recalibrated probability for the \NSoft{} flag targets and reached a mean flag AUC of \AucFlagSoft{} against \AucFlagHardSame{} with hard labels, with a squared correlation with the recalibrated probability of \RhoSqSoft{} against \RhoSqHardSame{}, so soft labels bring no gain to the estimator. The temporal condition gave Kumo Tabular small \TempCtxRows{} context rows from 2019 or from 2021 and scored test rows from each year. A context from the other report period changed the mean AUC by small amounts in both directions, and phone use shows the clearest effect, with an AUC on 2021 test rows of \TempPhoneMatchedTwentyOne{} from a 2021 context and \TempPhoneMismatchedTwentyOne{} from a 2019 context. The production comparison over all five folds used a gradient-boosted tree tuned per target by a random search of eight settings on an inner split of the training rows outside one fold, which favors the booster on the other folds.

\end{document}